\documentclass{aastex631}
\usepackage{tikz}
\usetikzlibrary{calc, angles, quotes, positioning, fit, matrix}
\usepackage{bm}
\usepackage{subcaption}
\usepackage[T1]{fontenc}
\usepackage{booktabs}
\usepackage{threeparttable}
\usepackage{multirow}
\usepackage{amsmath}
\usepackage[utf8]{inputenc}
\usepackage{makecell}
\usepackage{numprint}
\usepackage[ruled,vlined]{algorithm2e}
\usepackage{tabularx}

\begin{document}

 \title{CosIn: A Statistical-based Algorithm for Computation of Space-speed Time Delay in Pedestrian Motion}

\author{Jinghui Wang}

\affiliation{School of Safety Science and Emergency Management\\
Wuhan University of Technology\\
Wuhan, China}

\author{Wei Lv}
\altaffiliation{{\url{weil@whut.edu.cn} (W. Lv)}}
\affiliation{School of Safety Science and Emergency Management\\
Wuhan University of Technology\\
Wuhan, China}
\affiliation{China Research Center for Emergency Management\\
Wuhan University of Technology\\
Wuhan, China}

\author{Shuchao Cao}

\affiliation{School of Automotive and Traffic Engineering\\
Jiangsu University\\
Zhenjiang, China}

\author{Zhensheng Wang}

\affiliation{School of Artificial Intelligence\\
Beijing Normal University\\
Zhuhai, China}




\begin{abstract}
Precise assessment of Space-speed time delay (TD) is critical for distinguishing between anticipation and reaction behaviors within pedestrian motion. Besides, the TD scale is instrumental in the evaluation of potential collision tendency of the crowd, thereby providing essential quantitative metrics for assessing risk. In this consideration, this paper introduced the CosIn algorithm for evaluating TD during pedestrian motion, which includes both the CosIn-1 and CosIn-2 algorithms. CosIn-1 algorithm analytically calculates TD, replacing the numerical method of discrete cross-correlation, whereas the CosIn-2 algorithm estimates the TD from a statistical perspective. Specifically, the CosIn-1 algorithm addresses the precise computation of TD for individual pedestrians, while the CosIn-2 algorithm is employed for assessing TD at the crowd scale, concurrently addressing the imperative of real-time evaluation. Efficacy analyses of the CosIn-1 and CosIn-2 algorithms are conducted with data from single-file pedestrian experiments and crowd-crossing experiments, respectively. During this process, the discrete cross-correlation method was employed as a baseline to evaluate the performance of both algorithms, which demonstrated notable accuracy. This algorithm facilitate the precise evaluation of behavior patterns and collision tendency within crowds, thereby enabling us to understand the crowds dynamics from a new perspective.
\end{abstract}




\section*{Note}
Compared to the formally published version, this version has been revised to correct errors in Eq. \ref{1} (including Eq. \ref{31} and Fig. \ref{fig18}) and Eq. \ref{2}. A negative sign has been added to Eq. \ref{1}, and the order of variables in Eq. \ref{2} has been rearranged. I sincerely apologize for the errors and any inconvenience they may have caused.

\section{Introduction} 
\label{section1}

In traffic or single-file pedestrian motion \citep{cao2020analysis, tavana2024novel}, responses from pedestrians typically derive from the dynamic behavior exhibited by those ahead (stimulus–response behaviour) \citep{zheng2023parsimonious}. Consequently, within this scenario, a aggressive reaction behavior \citep{makridis2019estimating,kesting2008reaction} is observed among pedestrians (or drivers) as opposed to the conservative avoidance mechanism associated with anticipation behaviors \citep{karamouzas2014universal}. As a result, disturbances in the current unit’s state propagate upstream \citep{cordes2023single}. In a highly correlated space-speed response system, the difference between anticipation and reaction behaviors is straightforward, contingent upon speed and spatial variation over time. Reaction behavior is characterized by a delay in the alteration of speed relative to spatial variation, and denoted as reaction time.  In contrast, anticipation behavior is characterized by an pre-action in the alteration of speed concerning spatial variation, termed as anticipation time.

\begin{figure}[ht!]
\centering
\includegraphics[scale=0.5]{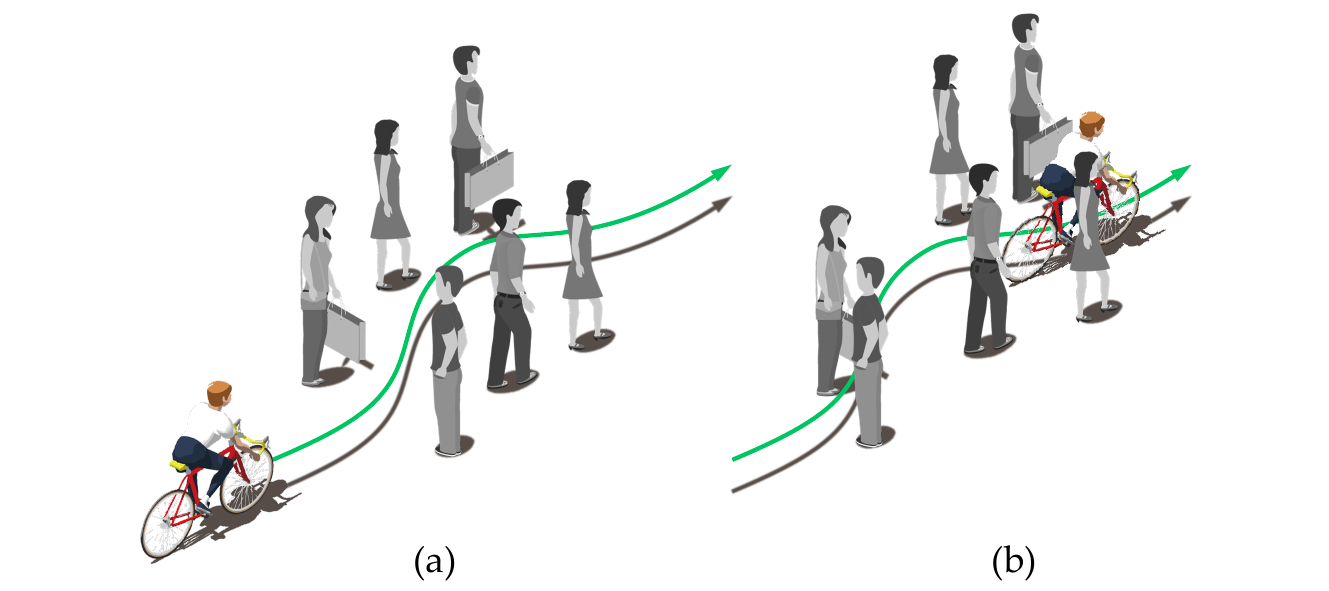}
\caption{Illustration of bicyclists navigating through a crowd.}
\label{fig1}
\end{figure}

Motion mechanism of pedestrians avoiding collisions through anticipation behaviors has been extensively observed. Such behaviors commonly manifest in human-involved traffic activities, including walking \citep{murakami2021mutual}, bicycle flow \citep{wangQ2023experimental}, vehicle flow \citep{treiber2006delays,chen2023modeling}, mixed traffic \citep{berge2024triangulating}, etc. Abundant research has demonstrated the effectiveness of similar mechanisms in preventing collisions within group dynamics 
 \citep{everett2021collision,zhang2021pedestrian}. Anticipation behaviors can be elucidated by the motion process of cyclists crossing through crowds. Deceleration is initiated by cyclists before approaching the crowd, even under negligible constraints, as depicted in Fig. \ref{fig1}(a). As cyclists are on the verge of leaving the crowd, they will accelerate proactively, as illustrated in Fig. \ref{fig1}(b). Similar motion mechanisms contributing to collision avoidance in collective dynamic \citep{gerlee2017impact,murakami2022spontaneous}. Such as the process of lane-changing, anticipation behavior plays a significant role in facilitating seamless lane changes \citep{chen2023modeling}. Within mixed traffic scenarios, anticipated behaviors serve as a foundation for the spontaneous order formation at crosswalks \citep{nirmale2024two,zheng2017model}. By endowing anticipation mechanisms, the efficiency of robots navigating through crowds is enhanced \citep{sathyamoorthy2020densecavoid}. Similarly, in the case of autonomous vehicles, pedestrians' anticipation motion are analyzed to facilitate trajectory prediction and collision avoidance 
 \citep{kotseruba2020they,rasouli2019pie}, to name a few. In general, utilizing algorithms based on anticipation effects or Time-To-Collision (TTC) 
 \citep{everett2021collision} significantly contributes to enhancing the resilience and reliability of robots or autonomous vehicles when confronted with complex environments.

\begin{figure}[ht!]
\centering
\includegraphics[scale=0.45]{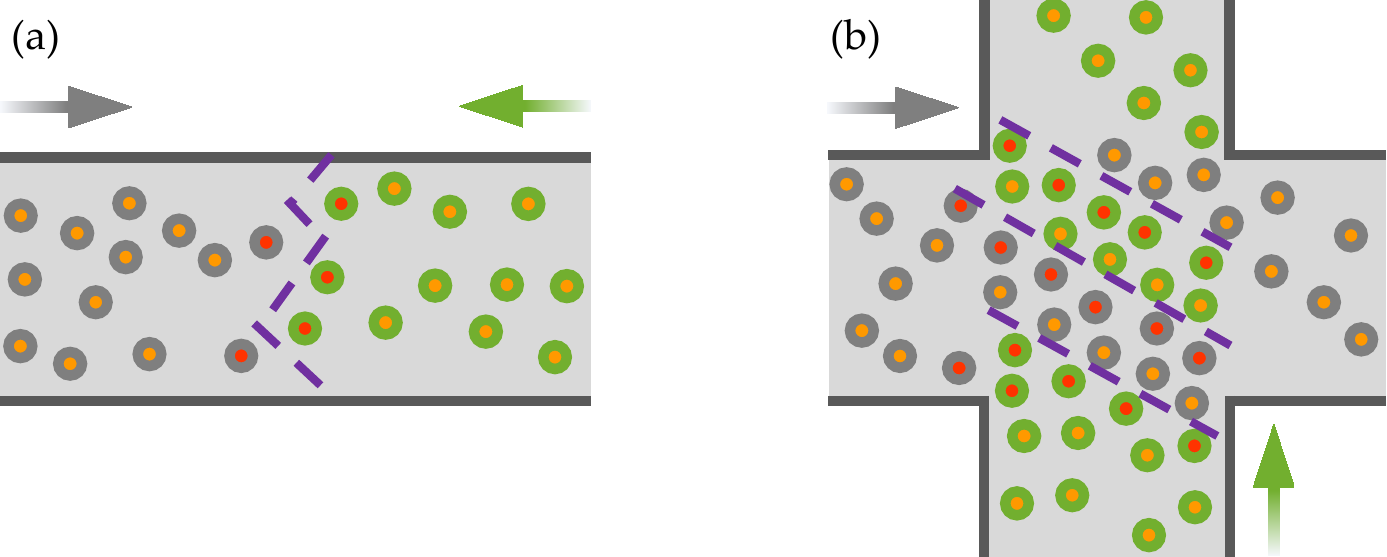}
\caption{Schematic diagram of pedestrian self-organization mechanisms. The green and gray outer circles represent pedestrians moving in different directions, while the red and yellow cores indicate pedestrians exhibiting anticipation or reaction behaviors, respectively. We conjecture that anticipation behaviors among pedestrians moving in diverse directions contribute to "phase separation" processes, as denoted by the purple dashed line.}
\label{fig2}
\end{figure}

In modeling, pedestrians have traditionally been considered as simple particles 
 \citep{helbing1995social} or social particles \citep{moussaid2010walking}, the interaction force among pedestrians is represented as a decay function of distance, constrains the velocity variation of pedestrians in differential time duration. However, the anticipation and reaction behaviors significantly contribute to the distinctions between pedestrian flow and granular flow 
 \citep{bonnemain2023pedestrians, cordes2021time, nicolas2019mechanical, zanlungo2011social, kleinmeier2020agent, yamamoto2019body, xiao2016pedestrian}. Collision avoidance is actively undertaken by pedestrians during motion \citep{chraibi2024elements}, a phenomenon prevalent in all human-involved motion activities. In contrast, granular flow manifests as free collision dynamics. From this perspective, pedestrian motion models based on anticipation and reaction behaviors can effectively prevent the phenomena of crowd freezing in dense crowds \citep{yi2023modeling,xu2024analysis}. More importantly, similar patterns of anticipation or reaction behaviors may constitute intrinsic rules of the pedestrian self-organization phenomena, like stripe \citep{zanlungo2023macroscopic} and lane formation \citep{feliciani2016empirical,bacik2023lane,murakami2021mutual}, etc. We conjecture that, anticipation behaviors contribute to the formation of leading fronts between the different directional flow (potential cooperative game induce "phase separation"), while reaction behaviors will lead to the formation of motion groups among the same directional flow \citep{wang2024CosForce}, as illustrated in Fig. \ref{fig2}.

Despite the intuitive and conceptually aligned nature of pedestrian motion, our understanding of these mechanisms remains limited, making it challenging to quantitatively assess their occurrence on both spatial and temporal scales. Specifically, questions such as "How far from obstacles do pedestrians initiate collision avoidance?" and "What is the time scale at which pedestrians avoid obstacles?" elude precise quantification. Insights into the spatial scale of anticipation behaviors (or reaction behaviors) during obstacle avoidance have been gained through a series of detour experiments 
 \citep{moussaid2009experimental,lv2013two}. But, these observational experiments were conducted in relatively simplistic scenarios, and analogous statistical methods 
 \citep{moussaid2009experimental} face challenges when applied in crowded environments. Due to the intricate dynamics of crowd, accurately determining the spatial scale becomes nearly impossible in high-density situations \citep{nicolas2019mechanical,sieben2023inside}. Consequently, focusing on temporal scales in statistics may offer a potential solution. We infer that the dynamism of crowds is governed by a mechanism in which an increase in environmental complexity is likely associated with a reduction in the TD scale of the crowd, ultimately leading to crowd dynamics resembling a simplified particle collision process \citep{zuriguel2011silo,patterson2017clogging}.

To this end, researchers have developed many effective quantitative methods for measuring inherent risks within crowds \citep{feliciani2018measurement,zanlungo2023pure,cordes2024dimensionless}. The methods based on multiple parameters, particularly vector parameters, still face challenges in real-time assessment from streaming data. Existing state-of-the-art approaches based on image recognition  \citep{alia2024novel} and monocular depth estimation (MDE) \citep{ranftl2020towards,miangoleh2021boosting,yang2024depth} encounter various hurdles in accurate evaluation. Similarly, mainstream methods relying on calibration, stereo sensors or cameras \citep{pouw2024high}, light detection and ranging (LiDAR) are also subject to limitations. Therefore, accurately quantifying the TD scale of pedestrian motion is effective for understanding and managing crowds, as it only requires simple spatial scalar data. More importantly, it's simple enough and follows intuition. Within a crowd, exists a linear relationship between pedestrian speed and their spatial constraints (such as headway, Nearest Neighbor Relative Distance (NNRD, see Sec. \ref{subsection2.1}) or other metrics), a relationship widely supported by empirical evidence, akin to the constant time-headway strategy observed in traffic (see Sec. \ref{subsection3.1}). By statistically analyzing pedestrian speed and spatial variation over time, rough estimates of TD can be obtained. In the context of sampled data, it is common practice to statistically analyze the TD corresponding to pairwise reference point. However, due to the influence of measurement error and stochastic noise, the results obtained are inherently inconsistent. Consequently, the challenge arises: how can we accurately undertake the statistical estimation of TD given this inherent variability?

Given the considerations, an algorithm for the quantitative calculation of TD is proposed in this paper, named CosIn. Specifically, the CosIn algorithm comprises two sub-algorithms: CosIn-1 and CosIn-2. CosIn-1 is employed for micro-level analysis, allowing for the precise computation for TD of individual pedestrian. It is noteworthy that this algorithm exhibits considerable complexity. Conversely, CosIn-2 is tailored for macro-level analysis, addressing scenarios involving crowds. By leveraging specific assumptions, this algorithm facilitates real-time estimation of TD within a designated region. 

The subsequent sections of this paper unfold as follows: in Section \ref{section2}, we provide definitions of anticipation and reaction behaviors. Additionally, we explore the manifestations of these mechanisms in various scenarios through observations from controlled experiments. Next, in Section \ref{section3}, the exposition is presented, delving into the background knowledge and theoretical underpinnings of the algorithms. The clarification of these aspects contributes to subsequent computational endeavors. Section \ref{section4} introduces the CosIn-1 algorithm, and validation ensues through empirical experiments involving single-file pedestrian motion. Section \ref{section5} delineates the CosIn-2 algorithm, accompanied by the validation procedures based on empirical data related to both single-file motion and crowd-cross scenarios. In this section, the errors and complexities of the CosIn-1 and CosIn-2 algorithms are compared, using the discrete cross-correlation method as the baseline. Finally, Section \ref{section6} encapsulates the conclusions drawn from the study.

\section{Anticipation and Reaction Behavior} \label{section2}

As discussion above, we illustrated the impact of anticipation and reaction behavior mechanisms on the research of crowd dynamics. Before arguing that this is indeed the case, let us try to sharpen the definition of the terminologies. What is the anticipation and reaction behavior? 

\subsection{Definitions} \label{subsection2.1}

In two-dimensional space, the interactions among pedestrians are multivariate rather than binary, and anisotropic rather than isotropic. Currently, there is no comprehensive method to fully present the spatial constraints of pedestrian motion. In this paper, we employ the metric of Nearest Neighbor Relative Distance (NNRD) 
 \citep{wang2023exploring} to quantitatively evaluate the spatial variations of pedestrian. The NNRD quantifies the relative distance between a reference pedestrian and their nearest neighbor within a defined Horizontal Field of Attention (HFA, constrained by the attentional eccentricity angle \(\phi\). Here, \(\phi\) denotes the angle between the boundary of sector-shaped HFA and the pedestrian's motion orientation).

Based on the evaluation of the NNRD \((\bm{d})\), we present the quantitative definitions of the time-related terminologies:

\textbf{Time-To-Collision (TTC)}: In two-dimensional space, the TTC of pedestrians can be computed as:

\begin{equation}\label{1}
\tau = 
\begin{cases}
\frac{-\sqrt{4r^2 - \|\bm{d}_{ij}\|^2 \sin^2 \theta} + \|\bm{d}_{ij}\| \cos \theta
}{\|\bm{v}_{ij}\|}, & \text{if }  \theta   \leq \arcsin \left( \frac{2r}{\| \bm{d}_{ij} \|} \right), \\[10pt]
\hfill \infty, \hfill & \text{otherwise}.
\end{cases}
\end{equation}

Here, \(\theta = \angle \left( {{\bm{d}_{ij}},{\bm{v}_{ij}}} \right) \ge 0\), \({\bm{d}_{ij}}\) denotes the head to head vector between pedestrian \(i\) and its nearest neighbor \(j\) in HFA (\(\phi=\pi \)), \( \bm{d}_{ij} = \bm{x}_{j} - \bm{x}_{i} \). \(\bm{v}_{ij}\) represents the relative velocity of pedestrian \( i \), considering pedestrian \( j \) as a stationary reference point, \( \bm{v}_{ij} =\bm{v}_i - \bm{v}_j \). \(r\) represents the equivalent radius of the pedestrian. For a detailed explanation of Eq.\ref{1}, please refer to Appx. \ref{Appendix B}. When \(\phi \to 0\), \({\tau }\) is denoted as the TTC for a single-file pedestrian or vehicle. In all cases, TTC adheres to the condition \(\tau \ge 0\).

The TTC-based metric exhibits the following limitations:

(1) The assessment of Time-to-Collision (TTC) is based on the current state to predict collision trends over a future period, assuming that the pedestrian's motion state (such as velocity, acceleration, angular velocity, etc.) remains constant over the evaluation period. This assumption does not align with the reality of pedestrian motion resembling a Markov process (operational level)—where the current state depends solely on the preceding moment, and future states exist probabilistically.

(2) The binary interaction of \(i\) and \(j\) typically relies on the nearest neighbor assumption (leader-follower interaction in string dynamics). This approach struggles to address multi-body interactions, making the evaluation of TTC in the two-dimensional space challenging.

\textbf{Space-speed Time Delay (TD)}: In non-free flow motion \((0<v<{{v}_\text{free}})\), the TD \(\left( \delta  \right)\) manifested as a time delay between space variation (such as headway, NNRD, or analogous measures) and speed variation in response to perturbation \(\left( \varepsilon  \right)\), performed as:

\begin{equation}\label{2}
\delta = t\left( {{\bm{d}}_{\varepsilon }} \right) -t\left( {{\bm{v}}_{\varepsilon }} \right).
\end{equation}

In this context, \(t\left( {{\bm{d}}_{\varepsilon }} \right)\) denotes the response moment of NNRD to perturbation \(\left( \varepsilon  \right)\), while \(t\left( {{\bm{v}}_{\varepsilon }} \right)\) denotes the response moment of speed to perturbation \(\left( \varepsilon  \right)\). When \(\delta >0\) holds, we characterize the pedestrian's behavior as anticipation behavior (the corresponding TD is also denoted as anticipation time). Conversely, when \(\delta <0\) is satisfied, we designate the pedestrian's behavior as reaction behavior (the corresponding TD be termed as reaction time) \citep{wang2024}.

It should be noted that, Eq.\ref{2} is specific to transient assessments. In analysis, a comprehensive assessment of pedestrian and crowd states can be achieved by adjusting the observation time window. When the time window is short, transient behavioral patterns can be observed, whereas expanding the time window enables the statistical analysis of long-term motion trends.

\textbf{Anticipation Time}: The pre-action duration \(\left( \delta >0 \right)\) during which pedestrians engage in anticipation maneuvers to avert collisions when encountering perturbation \(\varepsilon \).

\textbf{Reaction Time}: The post-action duration \(\left( \delta <0 \right)\) during which pedestrians engage in reaction maneuvers to avert collisions when encountering perturbation \(\varepsilon \).

\subsection{Anticipation and Reaction Mechanisms} \label{subsection2.2}

In this section, we evaluated pedestrian motion in several controlled experiments to explore the patterns of anticipation and reaction behavior. A total of three experiments were included in the observation: the binary interaction experiment \citep{murakami2022spontaneous}, the circle antipode experiment \citep{xiao2019investigation}, and the perpendicular crossflow experiment \citep{zanlungo2023macroscopic}. The pedestrian motion patterns in these experiments covered head-on motion, multi-directional motion, and cross motion.

\textbf{Binary Interaction Experiment}

\begin{figure}[ht!]
\centering
\includegraphics[scale=0.55]{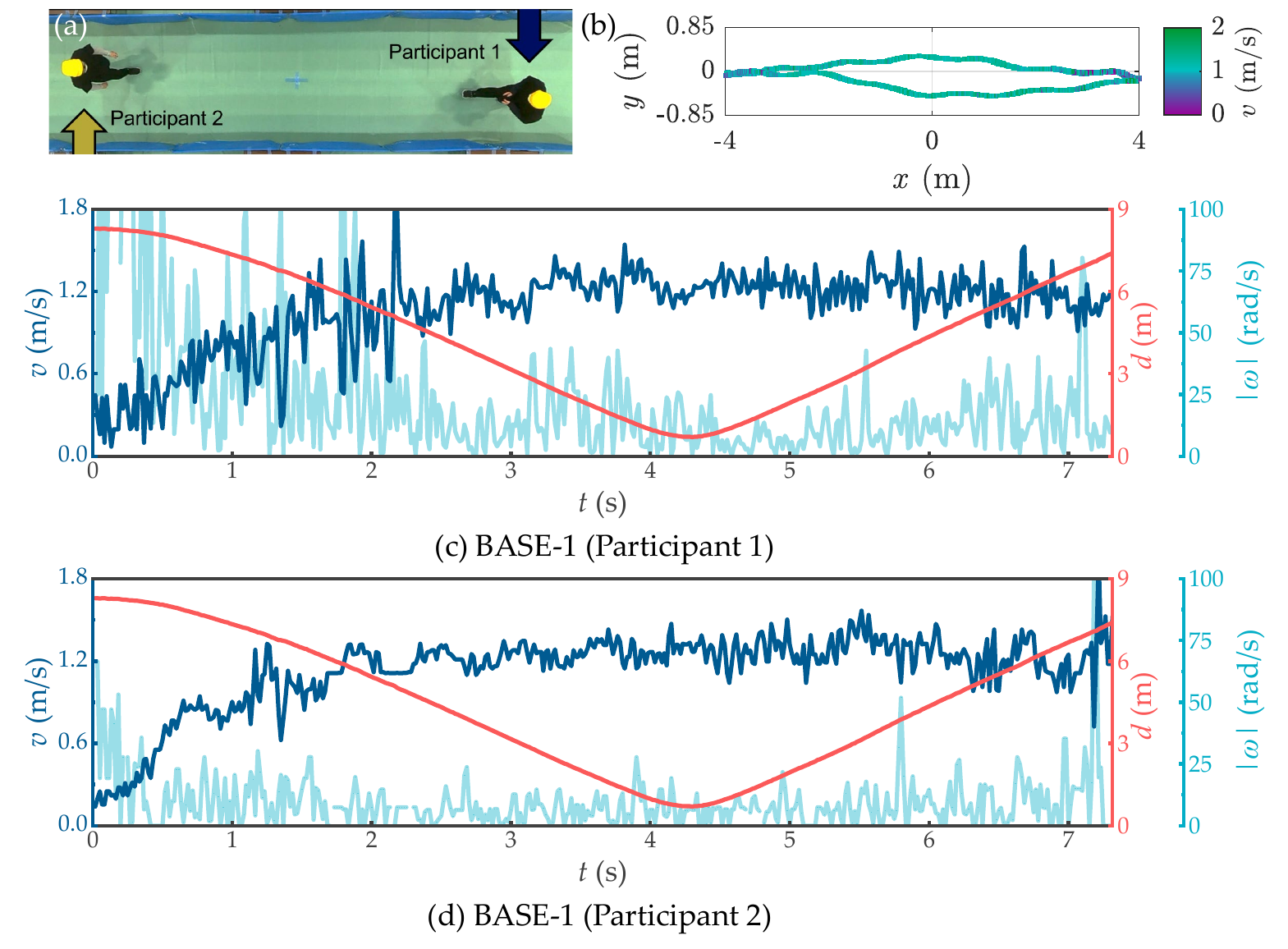}
\caption{Binary Interaction Experiment. (a) experimental snapshot, with the image sourced from \citep{murakami2022spontaneous}. (b) corresponding pedestrian trajectories. (c) and (d) illustrate the time series of motion state changes (speed, NNRD(\(\phi=\pi\)) and angular velocity) for participant 1 and participant 2,  respectively. The subtitles of the subfigures provide the corresponding data identifiers (BASE: no additional tasks for either participant).}
\label{fig3}
\end{figure}

Firstly, we observed the interactions between paired individuals, as depicted in the experimental snapshot in Fig. \ref{fig3}(a). This experiment investigated the interactions among pedestrians moving in head-on directions within a straight corridor 
 \citep{murakami2022spontaneous}. The data for this experiment was sourced from: \url{https://data.mendeley.com/datasets/rv89pk8cj2/1} ( Mendeley Data).

We extracted a set of experimental data for analysis (sampling frequency: 60 Hz), and the trajectories is shown in Fig. \ref{fig3}(b). Corresponding variations in the participants' speed, NNRD, and angular velocity are depicted in Figs. \ref{fig3}(c) and \ref{fig3}(d). In the baseline experiment (BASE), by observing the time series of NNRD, speed, and angular velocity of the participants, it is evident that participants initiated directional maneuvers (significant angular velocity) from the beginning position of the experiment (\({\bm{d}} \approx 8\)m) to avoid head-on collisions. These maneuvering strategies allowed participants to maintain their speed when encountering oncoming pedestrians. According to the analysis, lateral maneuvers by participants in the baseline experiment (BASE) spatially preceded those in the no mutual anticipation experiment (NMA), as illustrated in Fig. 1 and Fig. S1 in 
 \citet{murakami2022spontaneous}, which mechanism illustrated the anticipation mechanism in head-on motion.

In this scenario, the ample space allows pedestrians adopted the detour strategy rather than a deceleration strategy to avoid collisions. The motion state of the participants does not meet the prerequisites set forth in Eq. \ref{2} (i.e., non-free flow motion), nor does a synchronous relationship emerge between pedestrian speed and spatial constraints (\({\bm{d}}\)). However, the anticipation behavior of pedestrians in this setting is observable, as they engage in lateral maneuvers to execute avoidance even from considerable distances, where their motion in free flow is unimpeded.

\textbf{Circle Antipode Experiment}

\begin{figure}[ht!]
\centering
\includegraphics[scale=0.55]{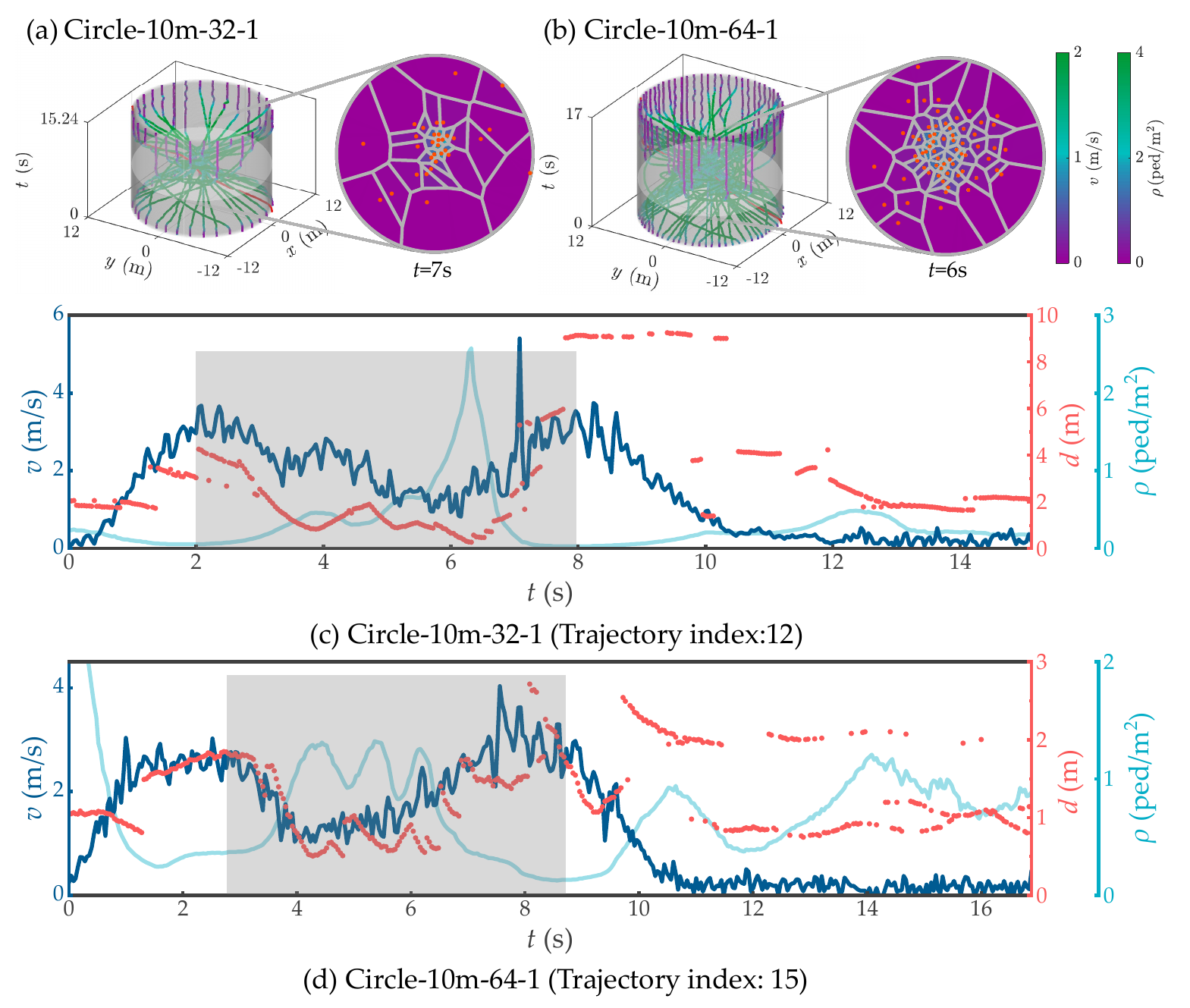}
\caption{Circle Antipode Experiment. (a) and (b) display the pedestrian trajectories and the corresponding distribution of local densities in two sets of experiments, respectively. The trajectories of the reference pedestrians highlighted in red. (c) and (d) illustrate the time series of changes in the motion states (speed, NNRD(\(\phi=\pi/2\)) and local density) of the reference pedestrians, respectively. The titles of the subfigures specify the corresponding data identifiers and trajectory indexes.}
\label{fig4}
\end{figure}

The circle antipode experiment investigates the interactions among pedestrian groups in multi-directional motion \citep{xiao2019investigation}. we extracted the data from two different sets of experiments (sampling frequency: 25 Hz), as depicted in Fig. \ref{fig4}. Figs. \ref{fig4}(a) and \ref{fig4}(b) illustrate the corresponding motion trajectories and the local density distribution based on the Voronoi method, respectively.

Figs. \ref{fig4}(c) and \ref{fig4}(d) present the time series of the reference pedestrians' state changes during the experiment. The gray areas in Figs. \ref{fig4}(c) and \ref{fig4}(d) demonstrate the mechanism of anticipation in group interactions, where variations in speed precede the NNRD. This phenomenon illustrated how pedestrians preemptively adjust their speed to maneuver through potential collisions during motion. The figures also presents the variations in corresponding local density. However, due to the weak negative correlation between local density and speed, discerning behavioral patterns becomes challenging.

Because of the potential noise present in the sampling data, the observations of the behavioral pattern depicted in Fig. \ref{fig4} remain ambiguous. In subsequent observations, we will analyze the noise-reduced data to gain more accurate insights.

\textbf{Perpendicular Crossflow Experiment}

\begin{figure}[ht!]
\centering
\includegraphics[scale=0.55]{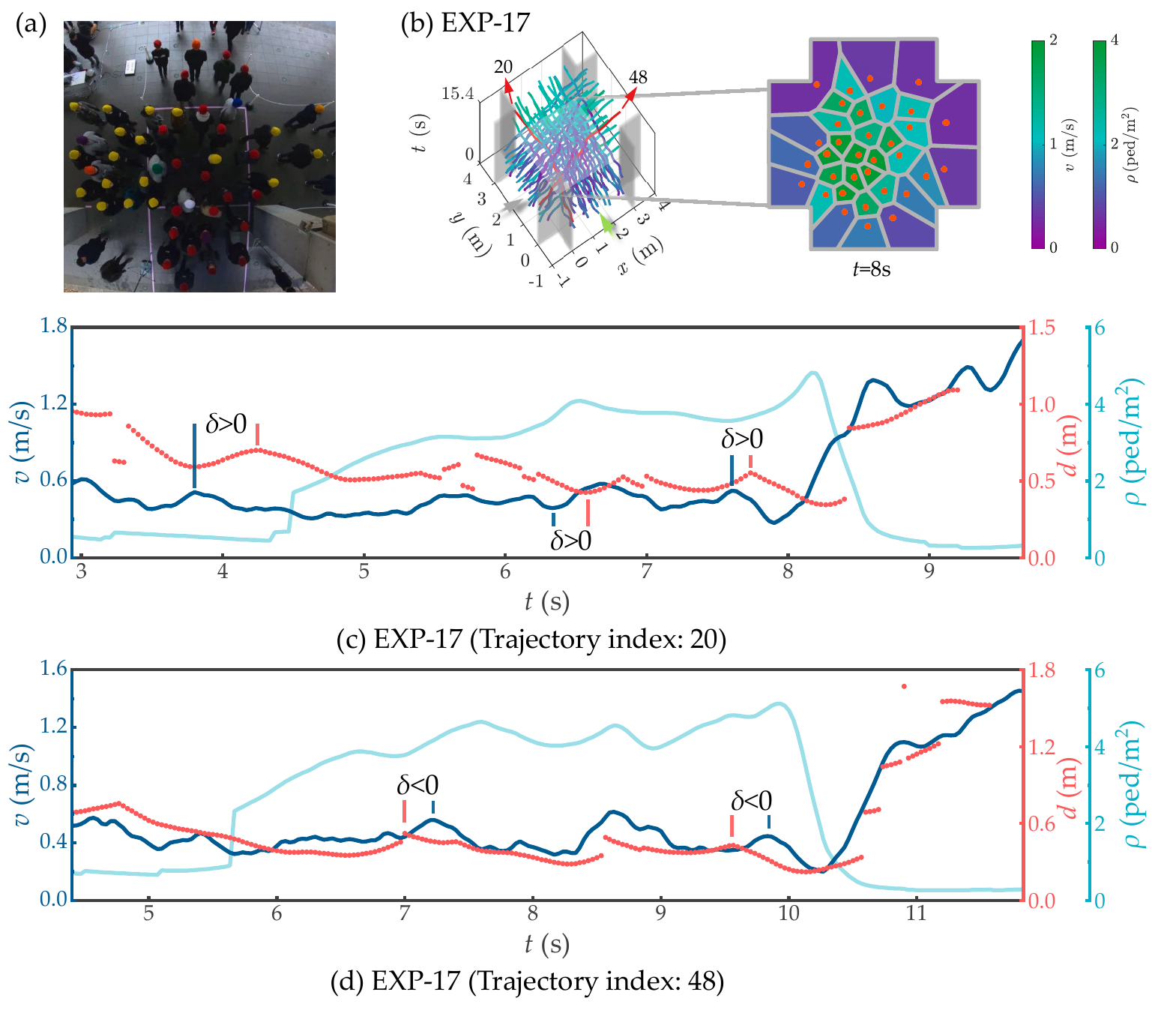}
\caption{Perpendicular Crossflow Experiment.  (a) experimental snapshot, with the image sourced from \citep{zanlungo2023macroscopic}. (b) corresponding pedestrian trajectories and distribution of local density, with the trajectories of the reference pedestrians highlighted in red. (c) and (d) illustrate the time series of motion state changes (speed, NNRD(\(\phi=\pi/2\)), and local density) of reference pedestrians, respectively. The subtitles of the subfigures provide the corresponding data identifiers and trajectory indexes.}
\label{fig5}
\end{figure}

The perpendicular crossflow experiment investigated pedestrian interactions in orthogonal motion to explore a typical self-organization phenomenon: stripe formation 
 \citep{zanlungo2023macroscopic}. A snapshot of the experiment is depicted in Fig. \ref{fig5}(a). The data for this experiment were sourced from: \url{https://ped.fz-juelich.de/da/doku.php?id=perpendicular_cross_flow} (Pedestrian Dynamics Data Archive), and all data have been subjected to smoothing processes.

We extracted two sets of pedestrian trajectories for analysis (sampling frequency: 30 Hz, indexed as 20 and 48). Fig. \ref{fig5}(b) presents a schematic diagram of the trajectories and local density distribution for the corresponding experiments, while Figs. \ref{fig5}(c) and \ref{fig5}(d) depict the time series of pedestrian motion states (speed, NNRD (\(\phi=\pi/2\)), and local density) of trajectories indexed 20 and 48, respectively. Fig. \ref{fig5}(c) reveals a clear anticipation mechanism, where speed changes precede spatial changes, i.e. \(\delta>0\). Conversely, Fig. \ref{fig5}(d) displays a typical reaction mechanism, with speed changes lagging behind spatial changes, i.e. \(\delta<0\) (similar pattern can be observed in Fig. 4 of 
 \citet{wang2021pedestrian}).

The observations from Figs. \ref{fig4} and \ref{fig5} suggest that spatial assessment methods based on local density may have certain limitations, as the changes in speed and variations in local density do not show a strong negative correlation. These findings highlight the limitations of evaluating pedestrian motion spaces using isotropic spatial measurement methods.

\subsection{Phase Diagram of TD and TTC} \label{subsection2.3}

\begin{figure}[ht!]
\centering
\includegraphics[scale=0.6]{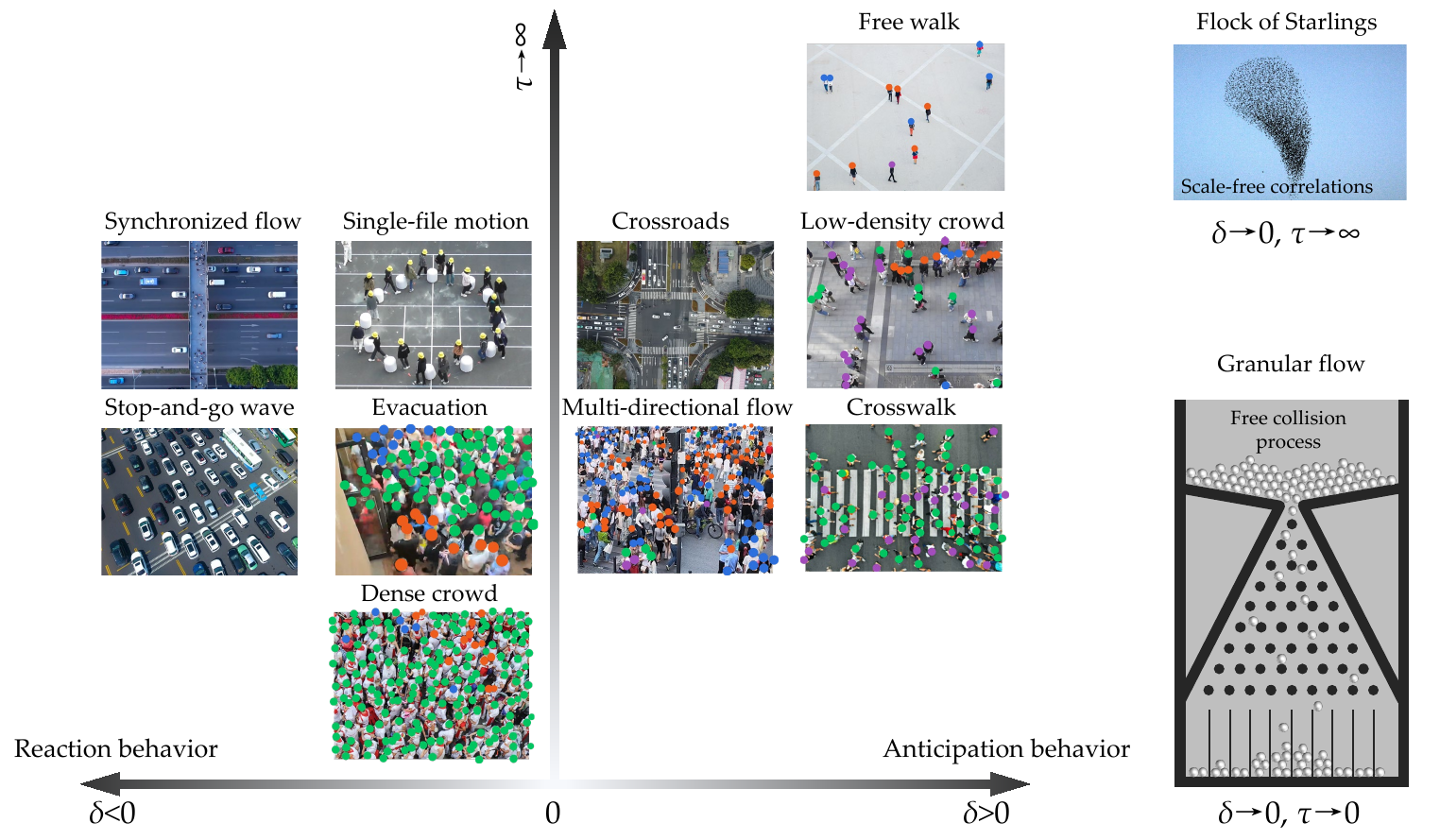}
\caption{Illustration of phase diagram pertain to  the TD and TTC. In the figures, dots colored green, purple, blue, and orange represent pedestrians moving leftward, rightward, downward, and upward, respectively. The image of flock of starlings sourced from \citet{couzin2018synchronization}. It is imperative to note that behavioral patterns within crowds are typically heterogeneous and subject to real-time variation. The depictions herein are intended merely as simplified examples and do not represent definitive results.}
\label{fig6}
\end{figure}

Based on the observations above, we investigated the anticipation and reaction behavior patterns within crowd dynamics. In this section, we constructed a phase diagram based on the two time constants of TD and TTC to categorize crowds, as depicted in Fig. \ref{fig6}.

The phase diagram comprises two dimensions: TTC (ranging from \(0 \to \infty\)) and TD (which includes both positive and negative values). A decrease in TTC indicates a tendency in pedestrian aggregation and corresponds to a heightened risk of collision. The TD scale divides the phase diagram into two quadrants (\(\delta < 0\) and \(\delta > 0\)), representing the dominance of reaction behavior and anticipation behavior, respectively.

We attribute string dynamics and high-density aggregation phenomena to the dominance of reaction behavior. In contrast, low-density group interactions, particularly in multi-directional flow dynamics (which may exhibit self-organizing tendencies), are classified as being dominated by anticipation behavior.

From this perspective, we posit that the increase in crowd aggregation (high density and dynamism)), leads to the reduction in TD and TTC scales. Consequently, as \(\delta \to 0\) and \(\tau \to 0\), the dynamics of the crowd converge to those of granular flow \citep{faure2015crowd}. Correspondingly, as \(\delta \to 0\) and \(\tau \to \infty\), the critical state emerges in the collective dynamics, characterized by scale-free correlations \citep{cavagna2010scale}. Undoubtedly, discussing criticality within the context of human crowds is almost a luxury.

\section{Theoretical Properties} \label{section3}

The perspectives outlined above highlight the pivotal role of pedestrians' anticipation behaviors and reaction behaviors within the framework of collective patterns. To further explore and validate these issues, there is a necessity for effective tools to facilitate objective and accurate measurements and assessments. In this section, we will commence with fundamental theories from the disciplines of pedestrian dynamics and signal processing, progressively building our methodology.

\subsection{Relationship between Space and Speed} \label{subsection3.1}

Ample empirical evidence attests that, in the case where pedestrians are neither in free flow nor at a complete standstill \((0<v<{{v}_\text{free}})\), exists a linear relationship between the available space in front of them and their speed 
 \citep{jelic2012properties,cao2020analysis,cordes2023single,wang2023exploring}. This relationship can be expressed by the following formula:

\begin{equation}\label{3}
d={{d}_{0}}+b\cdot v.
\end{equation}

Where, \(d\text{ }(\phi >0)\) denotes the NNRD, here \(\phi\) denotes attentional eccentricity angle \citep{wang2023exploring}. \({{d}_{0}}\) represents the upper bound of the NNRD for pedestrian keep standstill, and \(b\) is the regression coefficient. In Eq.\ref{3}, when \(\phi \to 0,\text{ }d\to h\), \(h\) represents the headway of pedestrian.

Therefore, an empirical formula for the time-headway to the pedestrian can be derived as follows:

\begin{equation}\label{4}
{{t}_{h}}=b+\frac{{{d}_{0}}}{v},\text{ }(\phi \to 0).
\end{equation}

Where, \({{t}_{h}}\) denotes the time-headway, and when \({{d}_{0}}=0\), signifies that pedestrians adopt the constant time-headway strategy.

\subsection{Orthogonality of Trigonometric Functions} \label{subsection3.2}

According to the orthogonality theorem of trigonometric functions, the following equations can be derived:

\begin{equation}\label{5}
\left\{ \begin{aligned}
  & \forall n,m\in \mathbb{Z}-\left\{ 0 \right\},\text{ }n=m,\text{ }\exists \int\limits_{T}{\sin nx\sin mx\text{ }dx=\int\limits_{T}{\cos nx\cos mx\text{ }dx=}\frac{T}{2}} \\ 
 & \forall n,m\in \mathbb{Z}-\left\{ 0 \right\},\text{ }n\ne m,\text{ }\exists \int\limits_{T}{\sin nx\sin mx\text{ }dx=\int\limits_{T}{\cos nx\cos mx\text{ }dx=}0} \\ 
 & \forall n,m\in \mathbb{Z}-\left\{ 0 \right\},\text{ }\exists \int\limits_{T}{\sin nx\cos mx\text{ }dx=0} 
\end{aligned} \right..
\end{equation}

Here, \(T\) denotes the common period of the integrated trigonometric functions. For any arbitrary sine or cosine functions \(A\) and \(B\), with periods \({{T}_{A}}\) and \({{T}_{B}}\) respectively, their common period is defined as: 

\begin{equation}\label{6}
{T}=\frac{\left| {{T}_{A}}\cdot {{T}_{B}} \right|}{GCD({{T}_{A}},{{T}_{B}})}.
\end{equation}

Where, \(GCD({{T}_{A}},{{T}_{B}})\) represents the greatest common divisor of \({{T}_{A}}\) and \({{T}_{B}}\).

\subsection{Correlation of Trigonometric Functions} \label{subsection3.3}

\begin{figure}[ht!]
\centering
\includegraphics[scale=0.85]{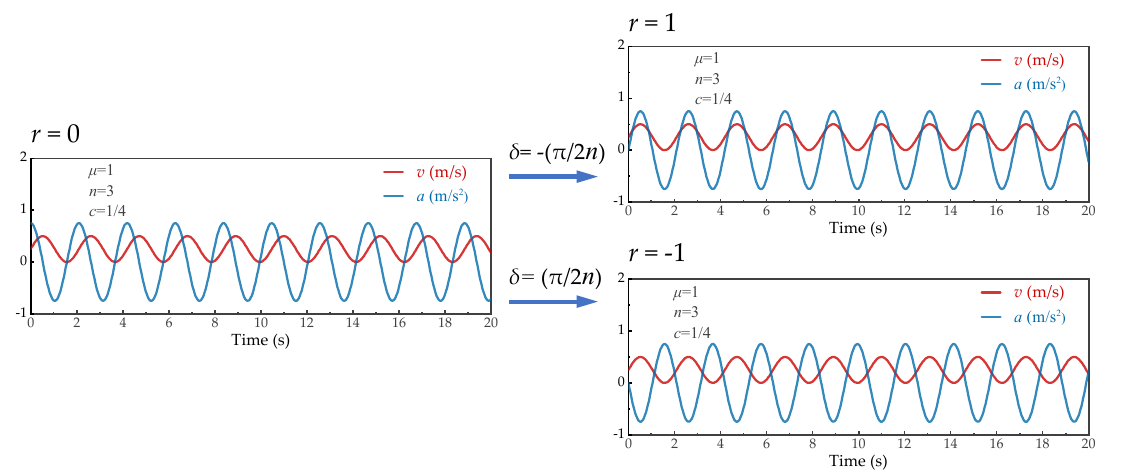}
\caption{Illustration of the relationship between the speed-time function and the corresponding acceleration-time shift function based on the sine function. The single-frequency signal (pure sine or cosine wave) can be regarded as a first-order Fourier series.
}
\label{fig7}
\end{figure}

For Fourier series with identical common periods, the correlation coefficient (Pearsons' \(r\)) between samples from two sets of series varies with the time shift (\(\delta\)). We constructed a speed-time history function, using the simplest sine function, denoted as \(v=\mu \cdot \sin \left( n\cdot t \right)+c\). Correspondingly, the acceleration time history function, denoted as \(a=\dot{v}=\mu \cdot n\cdot \cos \left( n\cdot t \right)\), is derived. By assigning parameters, we can depict its curves, as illustrated in Fig. \ref{fig7}. Here, let \(\delta \) denotes the time shift of the acceleration time history function.

According to Appx. \ref{Appendix C} and Eq. \ref{9}, the corresponding correlation coefficient and \textit{ZNCC} can be derived:

\begin{equation}\label{7}
r(\delta)=ZNCC(\delta)=\frac{{\int\limits_{t \in T} {\mu \sin (n \cdot t) \cdot \mu n\cos (n(t + \delta ))} dt}}{{\sqrt {\int\limits_{t \in T} {{{\left( {\mu \sin (n \cdot t)} \right)}^2}} dt} \sqrt {\int\limits_{t \in T} {{{\left( {\mu n\cos (n(t + \delta ))} \right)}^2}} dt} }}= - \sin \left( {n\cdot\delta } \right).
\end{equation}

When \(\delta =0\), the orthogonality of the sine and cosine functions implies that the correlation coefficient \(r\) between the continuous variables \(v\) and \(a\) is equal to zero. By shifting the acceleration-time function\(\left( \delta \ne 0 \right)\), the correlation coefficient between the variables \(v\) and \(a\) undergoes corresponding variations. In this specific case, when \(\delta =-\frac{\pi }{2\cdot n}\), \(r=1\), when \(\delta =\frac{\pi }{2\cdot n}\), \(r=-1\).

In the contemplation of a more common scenario, we address two Fourier series, denoted as \(F\) and \(G\), characterized by an identical period \(T\). Their respective expansions are expressed as follows:

\begin{equation}\label{8}
F=f(t)={{\alpha }_{0}}+\sum\nolimits_{n=1}^{\infty }{\left( {{\alpha }_{n}}\cos \left( \frac{2\cdot \pi \cdot n\cdot t}{T} \right)+{{\beta }_{n}}\sin \left( \frac{2\cdot \pi \cdot n\cdot t}{T} \right) \right)},
\end{equation}

\begin{equation}\label{9}
G=g(t)={{\mu }_{0}}+\sum\nolimits_{n=1}^{\infty }{\left( {{\mu }_{n}}\cos \left( \frac{2\cdot \pi \cdot n\cdot t}{T} \right)+{{\eta }_{n}}\sin \left( \frac{2\cdot \pi \cdot n\cdot t}{T} \right) \right)}.
\end{equation}

Here, \({{\alpha }_{0}}\) and \({{\mu }_{0}}\) correspond to the DC components of series \(F\) and \(G\), respectively, and \(\overline{F}={\alpha }_{0}\), \(\overline{G}={\mu }_{0}\). For series \(G\), when subjected to a time shift expressed as TD\(\left( \delta \right)\), the function \(G\left( \delta \right)\) can be constructed after translation:

\begin{equation}\label{10}
G\left( \delta  \right)=g\left( t+\delta  \right)={{\mu }_{0}}+\sum\nolimits_{n=1}^{\infty }{\left( {{\mu }_{n}}\cos \left( \frac{2\cdot \pi \cdot n\cdot \left( t+\delta  \right)}{T} \right)+{{\eta }_{n}}\sin \left( \frac{2\cdot \pi \cdot n\cdot \left( t+\delta  \right)}{T} \right) \right)}.
\end{equation}

According to the orthogonality theorem (refer to Sec. \ref{subsection3.2}), for the given sample data \(i=1,2,\cdot \cdot \cdot {\infty}\), the analytical results for the sample correlation coefficients arising from series \(F\) and \(G(\delta )\) can be derived as follows:

\begin{equation}\label{11}
\begin{aligned}
  & r(\delta )=ZNCC(\delta)=\frac{Cov(F,G(\delta ))}{{{\sigma }_{F}}{{\sigma }_{G(\delta )}}}=\frac{\sum\limits_{i=1}^{\infty}{\left({{F}_{i}}-\overline{F}\right) \left( {{G}_{i}}(\delta )-\overline{G(\delta )} \right)}}{\sqrt{\sum\limits_{i=1}^{\infty}{{{\left({{F}_{i}}-\overline{F}\right)}^{2}}\sum\limits_{i=1}^{\infty}{{{\left( {{G}_{i}}(\delta )-\overline{G(\delta )} \right)}^{2}}}}}} \\ 
 & =\frac{\int\limits_{T}{\left( \sum\nolimits_{n=1}^{\infty }{\left( {{\alpha }_{n}}\cos \left( \frac{2\cdot \pi \cdot n\cdot t}{T} \right)+{{\beta }_{n}}\sin \left( \frac{2\cdot \pi \cdot n\cdot t}{T} \right) \right)} \right)\cdot \left( \sum\nolimits_{n=1}^{\infty }{\left( {{\mu }_{n}}\cos \left( \frac{2\cdot \pi \cdot n\cdot (t+\delta )}{T} \right)+{{\eta }_{n}}\sin \left( \frac{2\cdot \pi \cdot n\cdot (t+\delta )}{T} \right) \right)} \right)dt}}{\sqrt{\int\limits_{T}{{{\left(\sum\nolimits_{n=1}^{\infty }{\left( {{\alpha }_{n}}\cos \left( \frac{2\cdot \pi \cdot n\cdot t}{T} \right)+{{\beta }_{n}}\sin \left( \frac{2\cdot \pi \cdot n\cdot t}{T} \right) \right)}\right)}^{2}}\text{ }dt}\cdot \int\limits_{T}{{{\left(\sum\nolimits_{n=1}^{\infty }{\left( {{\mu }_{n}}\cos \left( \frac{2\cdot \pi \cdot n\cdot (t+\delta )}{T} \right)+{{\eta }_{n}}\sin \left( \frac{2\cdot \pi \cdot n\cdot (t+\delta )}{T} \right) \right)}\right)}^{2}}\text{ }dt}}} \\ 
 & =\frac{\sum\limits_{n=1}^{\infty }{\left( {{\alpha }_{n}}\cdot {{\mu }_{n}}\cdot \cos \left( \frac{2\cdot \pi \cdot n\cdot \delta }{T} \right)+{{\alpha }_{n}}\cdot {{\eta }_{n}}\cdot \sin \left( \frac{2\cdot \pi \cdot n\cdot \delta }{T} \right)-{{\beta }_{n}}\cdot {{\mu }_{n}}\cdot \sin \left( \frac{2\cdot \pi \cdot n\cdot \delta }{T} \right)+{{\beta }_{n}}\cdot {{\eta }_{n}}\cdot \cos \left( \frac{2\cdot \pi \cdot n\cdot \delta }{T} \right) \right)}}{\sqrt{\sum\limits_{n=1}^{\infty }{\left( \alpha _{n}^{2}+\beta _{n}^{2} \right)}\cdot \sum\limits_{n=1}^{\infty }{\left( \mu _{n}^{2}+\eta _{n}^{2} \right)}}} \\ 
\end{aligned}.
\end{equation}

Similarly, the analytical results for the sample regression coefficients can be obtained as follows:

\begin{equation}\label{12}
\begin{aligned}
  & b(\delta )=r(\delta )\cdot \frac{{{\sigma }_{G(\delta )}}}{{{\sigma }_{F}}}=\frac{\sum\limits_{i=1}^{\infty}{\left( {{F}_{i}}-\overline{F} \right)\left( {{G}_{i}}(\delta )-\overline{G(\delta }) \right)}}{\sum\limits_{i=1}^{\infty}{{{\left( {{F}_{i}}-\overline{F} \right)}^{2}}}} \\ 
 & =\frac{\int\limits_{T}{\left( \sum\nolimits_{n=1}^{\infty }{\left( {{\alpha }_{n}}\cos \left( \frac{2\cdot \pi \cdot n\cdot t}{T} \right)+{{\beta }_{n}}\sin \left( \frac{2\cdot \pi \cdot n\cdot t}{T} \right) \right)} \right)\cdot \left( \sum\nolimits_{n=1}^{\infty }{\left( {{\mu }_{n}}\cos \left( \frac{2\cdot \pi \cdot n\cdot (t+\delta )}{T} \right)+{{\eta }_{n}}\sin \left( \frac{2\cdot \pi \cdot n\cdot (t+\delta )}{T} \right) \right)} \right)dt}}{\int\limits_{T}{{{\sum\nolimits_{n=1}^{\infty }{\left( {{\alpha }_{n}}\cos \left( \frac{2\cdot \pi \cdot n\cdot t}{T} \right)+{{\beta }_{n}}\sin \left( \frac{2\cdot \pi \cdot n\cdot t}{T} \right) \right)}}^{2}}\text{ }dt}} \\ 
 & =\frac{\sum\limits_{n=1}^{\infty }{\left( {{\alpha }_{n}}\cdot {{\mu }_{n}}\cdot \cos \left( \frac{2\cdot \pi \cdot n\cdot \delta }{T} \right)+{{\alpha }_{n}}\cdot {{\eta }_{n}}\cdot \sin \left( \frac{2\cdot \pi \cdot n\cdot \delta }{T} \right)-{{\beta }_{n}}\cdot {{\mu }_{n}}\cdot \sin \left( \frac{2\cdot \pi \cdot n\cdot \delta }{T} \right)+{{\beta }_{n}}\cdot {{\eta }_{n}}\cdot \cos \left( \frac{2\cdot \pi \cdot n\cdot \delta }{T} \right) \right)}}{\sum\limits_{n=1}^{\infty }{\left( \alpha _{n}^{2}+\beta _{n}^{2} \right)}} \\ 
\end{aligned}.
\end{equation}

We denote by the function \(\chi \) the quotient of the sample regression coefficients and the correlation coefficients. Consequently, we derive the functional expression of \(\chi \) with respect to the variable TD\(\left( \delta  \right)\):

\begin{equation}\label{13}
\begin{aligned}
  & \chi (\delta )=\frac{b(\delta )}{r(\delta )}=\frac{{{\sigma }_{G(\delta )}}}{{{\sigma }_{F}}}=\sqrt{\frac{\sum\limits_{i=1}^{\infty}{{{\left( {{G}_{i}}(\delta )-\overline{G(\delta )}\right)}^{2}}}}{\sum\limits_{i=1}^{\infty}{{{\left({{F}_{i}}-\overline{F}\right)}^{2}}}}} \\ 
 & =\sqrt{\frac{\int\limits_{T}{{{\sum\nolimits_{n=1}^{\infty }{\left( {{\mu }_{n}}\cos \left( \frac{2\cdot \pi \cdot n\cdot (t+\delta )}{T} \right)+{{\eta }_{n}}\sin \left( \frac{2\cdot \pi \cdot n\cdot (t+\delta )}{T} \right) \right)}}^{2}}\text{ }dt}}{\int\limits_{T}{{{\sum\nolimits_{n=1}^{\infty }{\left( {{\alpha }_{n}}\cos \left( \frac{2\cdot \pi \cdot n\cdot t}{T} \right)+{{\beta }_{n}}\sin \left( \frac{2\cdot \pi \cdot n\cdot t}{T} \right) \right)}}^{2}}\text{ }dt}}}=\sqrt{\frac{\sum\limits_{n=1}^{\infty }{\left( \mu _{n}^{2}+\eta _{n}^{2} \right)}}{\sum\limits_{n=1}^{\infty }{\left( \alpha _{n}^{2}+\beta _{n}^{2} \right)}}} \\ 
\end{aligned}.
\end{equation}

These results provide fundamental analytical relationships for Fourier series, and will significantly streamline the numerical computations of TD in the subsequent sections.

\section{Microscopic Analysis} \label{section4}

In this section, we undertake statistical analysis and computation based on individual pedestrian data, introducing the CosIn-1 algorithm. This algorithm relies on the Fourier approximation of speed and space data to compute the TD of pedestrian motion and can obtain an analytical solution for TD when the approximation error is controllable.

\subsection{CosIn-1 Algorithm} \label{subsection4.1}

For periodic functions that satisfy the Dirichlet convergence conditions, we can expand them into Fourier series. This allows us to easily construct Fourier series approximations as substitutes for the original function in expansion calculations. In the context of pedestrian motion scenarios, taking pedestrian speed and headway as examples, we can acquire temporal data on the speed and headway of pedestrians. Due to the inability to obtain truly continuous temporal data, the obtained sampled data essentially only reflects discrete states based on the sampling frequency \({{f}_{s}}\) and sampling time \({{T}_{s}}\), sampling size \(k={{f}_{s}}\cdot {{T}_{s}}\). Based on these, Fourier series of speed-time and headway-time functions are constructed. Since calculations are performed on discrete data, the process of Fourier expansion of the sampled data is Discrete Fourier Transformation (DFT). According to Nyquist's theorem, in order to avoid aliasing effects, the highest frequency component of the Fourier series is set to \({{f}_{\max }} = \frac{{{f}_{s}}}{2}\). The corresponding frequency resolution \( f_r \) is given by \( {f_{r}} = \frac{1}{T_s} \). consequently, the maximum expansion order of the Fourier series (\(N\)) should satisfy: \(N\le \frac{{{f_{\max }}}}{{{f_{r}}}} = \frac{k}{2}\). Appropriate expansion order selection facilitates the reduction of computational complexity while effectively controlling precision. In this study, we set \(N \leftarrow \lceil \frac{{k}}{10} \rceil\). So, the Fourier expansion expressions for the speed-time and headway-time functions are constructed in terms of series \(V\) and \(H\):

\begin{equation}\label{14}
V=v(t)={{\alpha }_{0}}+\sum\nolimits_{n=1}^{N}{\left( {{\alpha }_{n}}\cos \left( \frac{2\cdot \pi \cdot n\cdot t}{{{T}_{s}}} \right)+{{\beta }_{n}}\sin \left( \frac{2\cdot \pi \cdot n\cdot t}{{{T}_{s}}} \right) \right)},
\end{equation}

\begin{equation}\label{15}
H=h(t)={{\mu }_{0}}+\sum\nolimits_{n=1}^{N}{\left( {{\mu }_{n}}\cos \left( \frac{2\cdot \pi \cdot n\cdot t}{{{T}_{s}}} \right)+{{\eta }_{n}}\sin \left( \frac{2\cdot \pi \cdot n\cdot t}{{{T}_{s}}} \right) \right)}.
\end{equation}

As elucidated in Appx. \ref{Appendix C}, by performing a time shift on the Fourier series, we can formulate the series with respect to TD. Taking function \(H\) as an illustrative example, we construct its function \(H(\delta )\) with respect to TD \(\left( \delta  \right)\). This function is referred to as the shift function of headway-time.

\begin{equation}\label{16}
H(\delta )=h(t+\delta )={{\mu }_{0}}+\sum\nolimits_{n=1}^{N}{\left( {{\mu }_{n}}\cos \left( \frac{2\cdot \pi \cdot n\cdot \left( t+\delta  \right)}{{{T}_{s}}} \right)+{{\eta }_{n}}\sin \left( \frac{2\cdot \pi \cdot n\cdot \left( t+\delta  \right)}{{{T}_{s}}} \right) \right)}.
\end{equation}

According to Sec. \ref{subsection3.3}(see Eq. \ref{11}), it is established that through the application of time shift operations, the correlation coefficient between samples derived from functions \(V\) and \(H(\delta )\) can be computed. This correlation coefficient can be expressed as the function of \(\delta\):

\begin{equation}\label{17}
\begin{aligned}
  & r(\delta )=\frac{Cov(V,H(\delta ))}{{{\sigma }_{V}}{{\sigma }_{H(\delta )}}} \\ 
 & =\frac{\sum\limits_{n=1}^{N}{\left( {{\alpha }_{n}}\cdot {{\mu }_{n}}\cdot \cos \left( \frac{2\cdot \pi \cdot n\cdot \delta }{{{T}_{s}}} \right)+{{\alpha }_{n}}\cdot {{\eta }_{n}}\cdot \sin \left( \frac{2\cdot \pi \cdot n\cdot \delta }{{{T}_{s}}} \right)-{{\beta }_{n}}\cdot {{\mu }_{n}}\cdot \sin \left( \frac{2\cdot \pi \cdot n\cdot \delta }{{{T}_{s}}} \right)+{{\beta }_{n}}\cdot {{\eta }_{n}}\cdot \cos \left( \frac{2\cdot \pi \cdot n\cdot \delta }{{{T}_{s}}} \right) \right)}}{\sqrt{\sum\limits_{n=1}^{N}{\left( \alpha _{n}^{2}+\beta _{n}^{2} \right)}\cdot \sum\limits_{n=1}^{N}{\left( \mu _{n}^{2}+\eta _{n}^{2} \right)}}} \\ 
\end{aligned}.
\end{equation}

Now, we have constructed the correlation function \(r(\delta )\) between the samples of speed-time function and the time shift function of headway. 
The potential relationship between pedestrian speed and headway implying that a critical point can be achieved by adjusting the time shift function of headway, i.e., modifying the value of \(\delta \). At the critical point, the correlation between speed and headway is maximized (Theoretical basis is presented in Appx. \ref{Appendix C}): \(\delta_A = \mathop {\arg \max }\limits_{t \in \mathbb{R}} \left( r \left( \delta \right) \right)\) . The corresponding value \(({\delta }_{A})\) serves as the precise solution of TD. Consequently, through the differentiation of the correlation function, we can compute the exact solution. Here, we define the function as \(e\left( \delta  \right)\):

\begin{equation}\label{18}
\begin{aligned}
  & e\left( \delta  \right)=\frac{d\left( r\left( \delta  \right) \right)}{d\delta }= \\ 
 & \frac{\frac{2\cdot \pi }{{{T}_{s}}}\cdot \sum\limits_{n=1}^{N}{\left( n\cdot \left( {{\alpha }_{n}}\cdot {{\eta }_{n}}\cdot \cos \left( \frac{2\cdot \pi \cdot n\cdot \delta }{{{T}_{s}}} \right)-{{\alpha }_{n}}\cdot {{\mu }_{n}}\cdot \sin \left( \frac{2\cdot \pi \cdot n\cdot \delta }{{{T}_{s}}} \right)-{{\beta }_{n}}\cdot {{\mu }_{n}}\cdot \cos \left( \frac{2\cdot \pi \cdot n\cdot \delta }{{{T}_{s}}} \right)-{{\beta }_{n}}\cdot {{\eta }_{n}}\cdot \sin \left( \frac{2\cdot \pi \cdot n\cdot \delta }{{{T}_{s}}} \right) \right) \right)}}{\sqrt{\sum\limits_{n=1}^{N}{\left( \alpha _{n}^{2}+\beta _{n}^{2} \right)}\cdot \sum\limits_{n=1}^{N}{\left( \mu _{n}^{2}+\eta _{n}^{2} \right)}}} \\ 
\end{aligned}.
\end{equation}

By solving the algebraic equation \(e\left( \delta  \right)=0\), the TD \({({\delta }_{A})}\) of pedestrian motion can be obtained. Based on the numerical results from TD, we can thus determine the pedestrian behavior patterns: reaction behavior for \( {\delta }_{A} < 0 \) or anticipation behavior for \( {\delta }_{A} > 0 \).

We have designated the algorithm as CosIn-1, Algo. \ref{CosIn-1} provided the pseudocode representation. The algorithm under consideration does not employ loop structures, rendering its computational process relatively straightforward. Its complexity primarily stems from two components: DFT and solve the algebraic equations. In the subsequent section, empirical validation has been conducted to assess the applicability of the algorithm.

\begin{algorithm}
  \caption{CosIn-1 Algorithm}\label{CosIn-1}
  \SetAlgoLined
  \KwData{Pedestrian headway-time series $H$ and speed-time series $V$}

  \textbf{Input:} $H = \{ h(t_1), h(t_2), \ldots, h(t_{{k}}) \}$\\
  \text{  }{  }{  }{  }{  }{  }{  }{  }{  }{  }{  }{  } $V = \{ v(t_1), v(t_2), \ldots, v(t_{{k}}) \}$\\
  \textbf{Output:} $\delta_A$\\
  \textbf{Step 1:} Determine the order of the Fourier series:\\
  $N \leftarrow \lceil \frac{{k}}{10} \rceil$\\

  \textbf{Step 2:} Calculate Fourier series using discrete Fourier transform:\\
  $V \leftarrow \alpha_0 + \sum_{n=1}^{N} \left( \alpha_n \cos\left( \frac{2\pi n t}{T_s} \right) + \beta_n \sin\left( \frac{2\pi n t}{T_s} \right) \right)$\\
  $H \leftarrow \mu_0 + \sum_{n=1}^{N} \left( \mu_n \cos\left( \frac{2\pi n t}{T_s} \right) + \eta_n \sin\left( \frac{2\pi n t}{T_s} \right) \right)$\\

  \textbf{Step 3:} Construct the time shift function of headway:\\
  $H(\delta) \leftarrow \mu_0 + \sum_{n=1}^{N} \left( \mu_n \cos\left( \frac{2\pi n (t + \delta)}{T_s} \right) + \eta_n \sin\left( \frac{2\pi n (t + \delta)}{T_s} \right) \right)$\\

  \textbf{Step 4:} Calculate correlation coefficient:\\
  $r(\delta) \leftarrow \text{corr}(V, H(\delta))$\\

  \textbf{Step 5:} Differentiate $r(\delta)$:\\
  $e(\delta) \leftarrow \frac{d(r(\delta))}{d\delta}$\\

  \textbf{Step 6:} Construct differential algebraic equation $e(\delta) = 0$ and solve:\\
  $\delta_i, e(\delta_i) \leftarrow \text{Solve Equation:} \: e(\delta) = 0$\\

  \textbf{Step 7:} Find critical time shift (TD):\\
  $\delta_A = \mathop {\arg \max }\limits_{t \in \mathbb{R}} \left( r \left( \delta \right) \right)$\\

\end{algorithm}

\subsection{Case Study} \label{subsection4.2}

In this section, we conduct empirical validation of the CosIn-1 algorithm proposed in the preceding text. The data were derived from the experimental dataset involving a single-file pedestrian motion, with the experimental scenario depicted in Appx. \ref{Appendix A} (see Fig. \ref{fig17}). Detailed experimental procedures can be found by \citet{cao2019dynamic}. During the experiment, pedestrians were equipped with three different plastic glasses with lenses of limited light transmission (LT) values, namely LT=0\% (no visibility), LT=0.1\%, and LT=0.3\%. Data for pedestrian speed and headway were specifically extracted from a participants size of 30 (sampling frequency: 25 Hz, sampling durations: 14.56 s for LT = 0\%, 10.76 s for LT = 0.1\%, and 7.20 s for LT = 0.3\%).

\begin{figure}[ht!]
\centering
\includegraphics[scale=0.55]{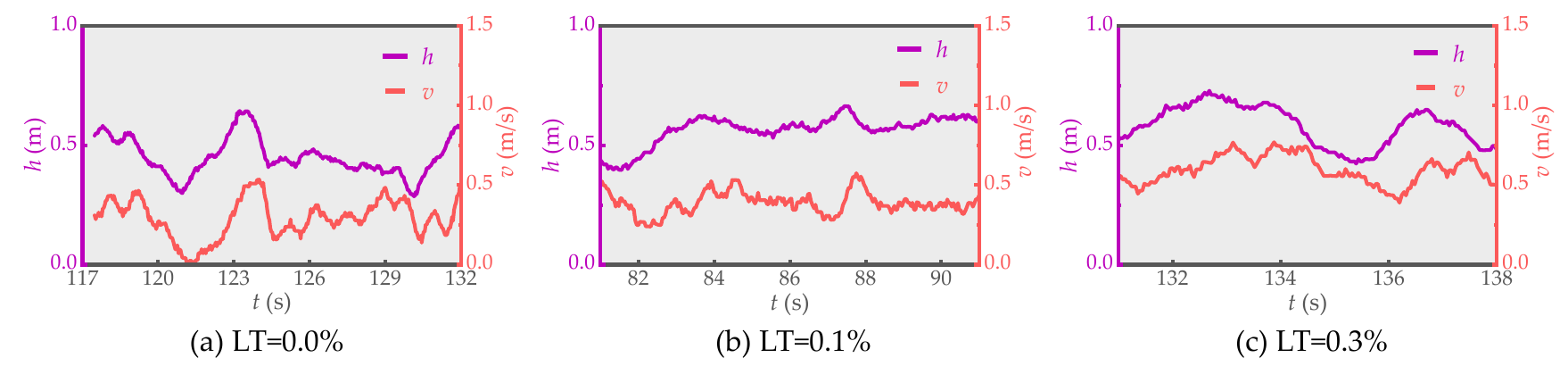}
\captionsetup{singlelinecheck=false, justification=centering}
\caption{Temporal data of speed profiles and headway profiles for individual pedestrians (participants size: 30), the sampling durations were 14.56 s, 10.76 s, and 7.20 s, respectively.}
\label{fig8}
\end{figure}

Fig. \ref{fig8} illustrates the temporal data of individual pedestrians' speed and headway across three experimental conditions. It is discernible from the graph that both speed and headway exhibit analogous temporal trends. Scrutinizing the temporal arrangement of the curves reveals a discernible time delay in speed variation compared to headway. Based on the discretely sampled data, the Fourier series of the speed-time function \(V=v(t)\) and the headway-time function \(H=h(t)\) can be computed. The sampling duration \({{T}_{s}}\) for both the speed data and the headway data is consistent, consequently, the common period \({{T}_{s}}\) of the speed-time function and the headway-time function can be established.

\begin{figure}[ht!]
\centering
\includegraphics[scale=0.78]{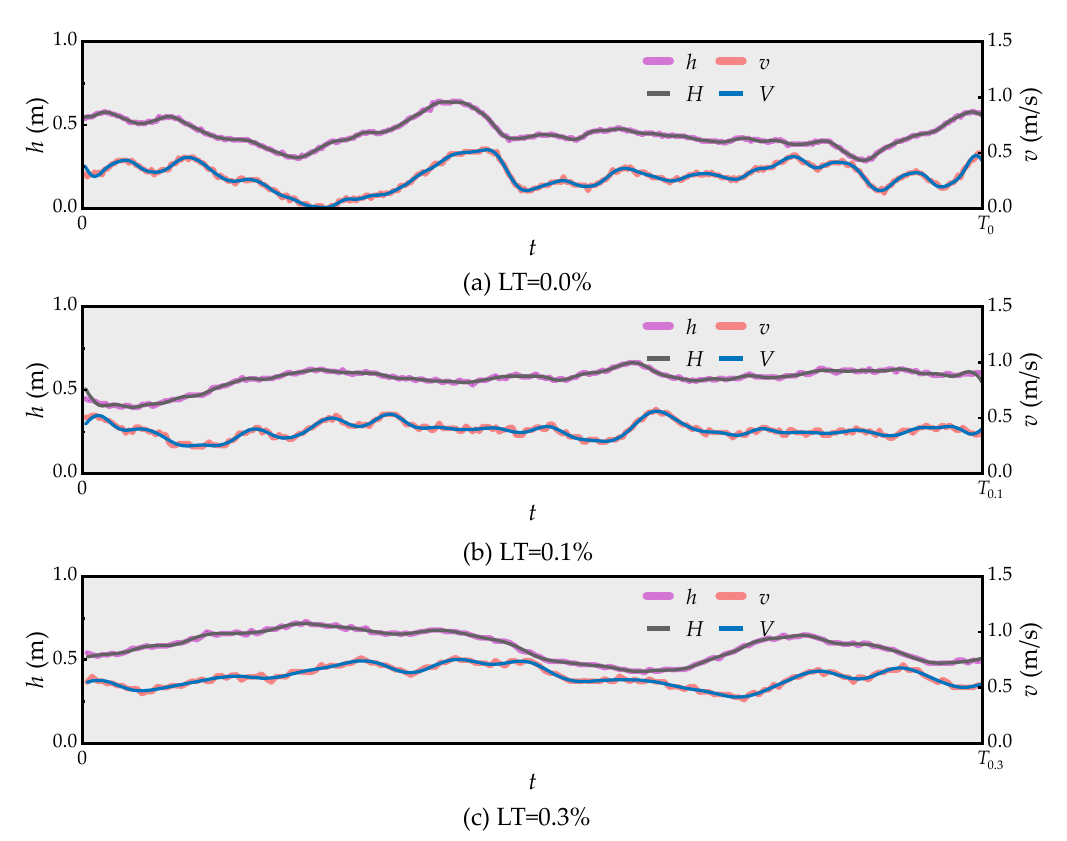}
\caption{Temporal data pertaining to pedestrian motion states and their corresponding Fourier series. Corresponding to LT = 0.0\%, LT = 0.1\%, and LT = 0.3\%, the orders of the Fourier series are 37, 27, and 19, respectively. For the Fourier coefficients, see Appx. \ref{Appendix E}.
}
\label{fig9}
\end{figure}

\begin{figure}[ht!]
\centering
\includegraphics[scale=0.6]{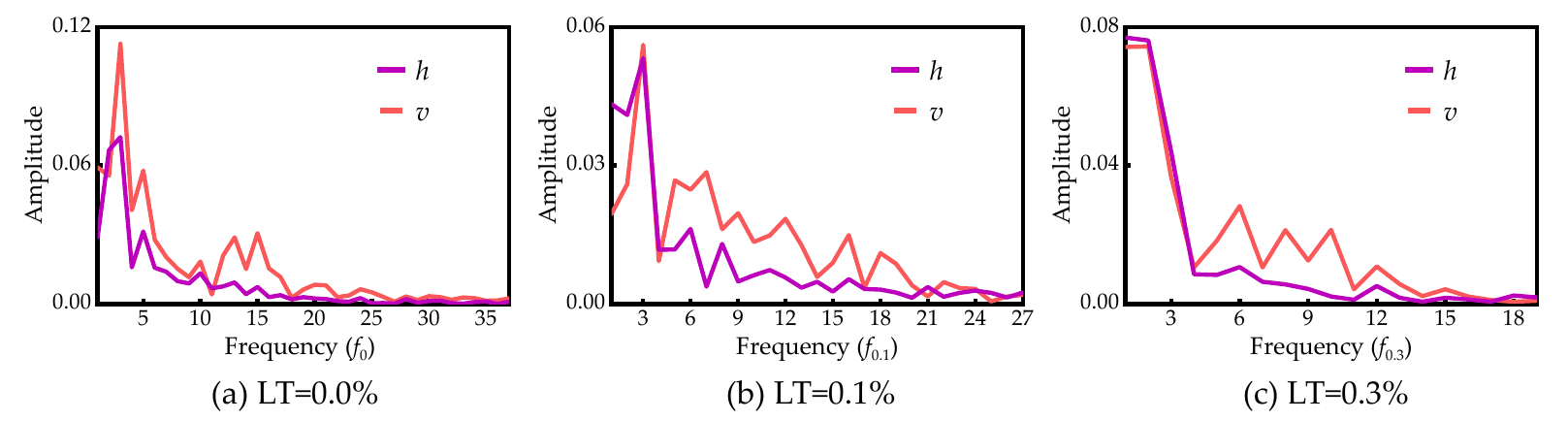}
\caption{Spectrogram corresponding to the Fourier series of three experimental sets. Here, \(f_0\), \(f_{0.1}\), and \(f_{0.3}\) correspond to the frequency resolutions of each data set, respectively, satisfying \(f_0 = {T_0}^{-1}\), \(f_{0.1} = {T_{0.1}}^{-1}\), and \(f_{0.3} ={T_{0.3}}^{-1}\). For the Fourier coefficients, see Appx. \ref{Appendix E}.}
\label{fig10}
\end{figure}

The original data and the Fourier series of three sets of experiments are depicted in Fig. \ref{fig9}. It is evident that the Fourier series provides a satisfactory approximation to the original data. Therefore, in subsequent sections, the Fourier series of speed function \(V=v(t)\) and the headway function \(H=h(t)\) are employed as substitutes for the original data in the computation. Fig. \ref{fig10} presents the spectrograms of the Fourier transforms applied to the speed and headway data in the three sets. The spectrograms for both speed and headway data exhibit remarkable proximity across the three experimental sets, indicating a consistent trend in the variations. Appx.\ref{Appendix E} provides the Fourier coefficients data of the three sets.

\begin{figure}[ht!]
\centering
\includegraphics[scale=0.55]{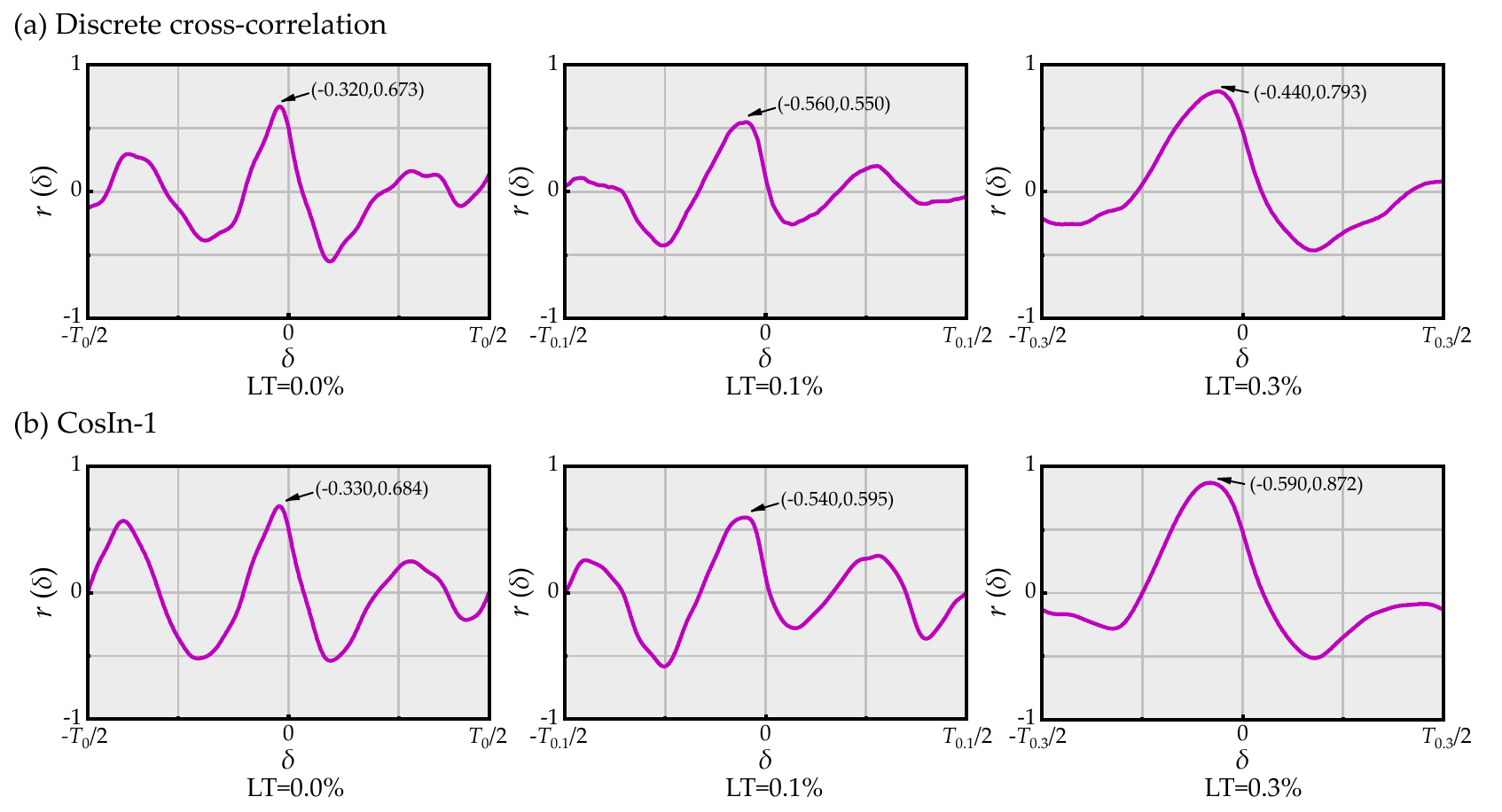}
\caption{Plots of the \(r(\delta )\) in three experimental groups, with numerical solutions of \({{\delta }_{A}}\) indicated at the arrows.}
\label{fig11}
\end{figure}

To facilitate a quantitative error evaluation of the CosIn-1 algorithm, the discrete cross-correlation was used as the baseline method, and the \(r(\delta)\) curves were compared. The graphical representation of \(r(\delta)\) is shown in Fig. \ref{fig11}, and the computed results of the CosIn-1 algorithm are presented in Tab. \ref{table1}. As depicted in the figure, the \(r(\delta)\) curves of both the discrete cross-correlation method and the CosIn-1 method are relative similar. Because of the use of Fourier series for substitution calculations in the CosIn-1 algorithm, the resulting \(r(\delta)\) curves is smoother. Based on the results in Tab.\ref{table1}, all TD values are negative, indicating that the reference participants consistently exhibited reaction behaviors in single-file motion. Comparative analysis of the computational results reveals that as the LT decreases, the temporal scale of reaction behaviors among reference pedestrians declines. This suggests that, within the given experimental setting, visual limitations may cause a contraction in the temporal scale of pedestrian reaction behaviors (i.e., a reduction in the absolute value of TD). It should be noted that our analysis was confined to the motion dynamics of a single reference participant in each experimental set over a period of several seconds, hence the computational results are impossible to generalize into empirical conclusions.

\begin{table}
  \centering
  \caption{The computational results of \({{\delta }_{A}}\).}
  \begin{tabular}{>{\centering\arraybackslash}p{3.5cm}>{\centering\arraybackslash}p{2cm}>{\centering\arraybackslash}p{2cm}>{\centering\arraybackslash}p{2cm}}
    \toprule
    \textbf{Experiment Index} & \textbf{LT=0\%} & \textbf{LT=0.1\%} & \textbf{LT=0.3\%} \\
    \midrule
    ${\delta}_{A}$ (s) & -0.329866 & -0.540323 & -0.590063 \\
    \bottomrule
  \end{tabular}
  \label{table1}
\end{table}

\section{Macroscopic Statistics}\label{section5}

The CosIn-1 algorithm, as elaborated in the last section, presents a methodology for accurately calculating TD within individual pedestrians. Precise TD computation is achieved by this algorithm. However, it mainly involves the DFT calculation, utilizing the Fast Fourier Transform (FFT) algorithm, with a time complexity of \( O(n \log n) \). Given the crowd context, an escalation in the number of sampled individuals, sampling duration, or sampling frequency leads to a power-law expansion of sample data. In crowd management, beyond ensuring the accurate calculation of TD, a paramount consideration is how to expedite TD computation when confronted with substantial data. The dynamic and transient nature of crowds underscores the importance of promptly, or even real-time, providing feedback—an essential aspect in ensuring safety. In this consideration, the CosIn-2 algorithm is proposed, which, based on certain assumptions, facilitates the rapid estimation of TD within crowds.

\subsection{CosIn-2 Algorithm}\label{subsection5.1}

By examining the spectrograms as illustrated in Fig. \ref{fig10}, it is discerned that the speed and headway (space-related) data of pedestrians is primarily composed of dominant frequencies (the fundamental frequency represents the lowest frequency component in the Fourier series, and the dominant frequency corresponds to the frequency associated with the highest amplitude). Consequently, we formulate the following assumptions for approximate calculations:

\textbf{Assumptions}:

(1) Representation of the temporal dynamics of speed through sine or cosine functions for approximation.

(2) Uniformity in the waveform of pedestrian dynamics within the region of interest (ROI), indicating that pedestrians within this area share a common waveform characterized by frequency. This can be regarded as a form of averaging operation to a certain extent.

Based on the assumptions above, we employ sine functions to characterize the speed-time function of pedestrians, represented as follows:

\begin{equation}\label{19}
V=v(t)={{\mu }_{c1}}\cdot \sin ({{n}_{c}}\cdot t)+{{c}_{1}}.
\end{equation}

Accordingly, the corresponding acceleration-time function can be established:

\begin{equation}\label{20}
A=a(t)=\dot{v}(t)={{\mu }_{c1}}\cdot {{n}_{c}}\cdot \cos ({{n}_{c}}\cdot t).
\end{equation}

Similarly, a shift parameter \(\lambda\) is introduced to formulate the time-shift function of acceleration.

\begin{equation}\label{21}
A(\lambda )=a(t+\lambda )={{\mu }_{c1}}\cdot {{n}_{c}}\cdot \cos \left( {{n}_{c}}\cdot \left( t+\lambda  \right) \right).
\end{equation}

The functions \(V\), \(A\), and \(A(\lambda )\) share a common frequency \({f}_{c}\), with corresponding periods \({{T}_{c}}\) and frequencies \({f}_{c}\) given as follow:

\begin{equation}\label{22}
{{f}_{c}}=\frac{1}{{{T}_{c}}}=\frac{{{n}_{c}}}{2\pi }.
\end{equation}

Where \({n}_{c}\), originating from Eq.\ref{19}, is represented as the frequency factor.

Assuming the pedestrian count within the ROI is denoted as \(m\), the individual sampling size is represented by \(k\), and the total sampling size denoted as \(S=m\cdot k\). According to Eq.\ref{11}, the sample correlation coefficients of the corresponding speed-time history function \(V\) and the time shift function of  acceleration \(A(\lambda )\) are determined:

\begin{equation}\label{23}
\begin{aligned}
  & {{r}_{a-v}}(\lambda )=\frac{Cov(A(\lambda ),V)}{{{\sigma }_{A(\lambda )}}{{\sigma }_{V}}} \\ 
 & =\frac{\sum\limits_{i \in S}{\left({{A}_{i}}(\lambda )-\overline{A}\right) \left( {{V}_{i}}-\overline{V} \right)}}{\sqrt{\sum\limits_{i \in S}{{{\left({{A}_{i}}(\lambda )-\overline{A}\right)}^{2}}\cdot\sum\limits_{i \in S}{{{\left( {{V}_{i}}-\overline{V} \right)}^{2}}}}}}\simeq\frac{\int\limits_{{T}_{c}}{\cos \left( {{n}_{c}}\cdot \left( t+\lambda  \right) \right)\cdot \sin \left( {{n}_{c}}\cdot t \right)\text{ }dt}}{\sqrt{\int\limits_{{T}_{c}}{{{\cos }^{2}}({{n}_{c}}\cdot (t+\lambda ))dt}\cdot \int\limits_{{T}_{c}}{{{\sin }^{2}}({n}_{c}\cdot t)dt}}}=-\sin \left( {{n}_{c}}\cdot \lambda  \right) \\ 
\end{aligned}.
\end{equation}

And, the corresponding function of regression coefficient \({b}_{a-v}\left( \lambda  \right)\) is:

\begin{equation}\label{24}
\begin{aligned}
  & {{b}_{a-v}}\left( \lambda  \right)={{r}_{a-v}}\left( \lambda  \right)\cdot \frac{{{\sigma }_{A(\lambda )}}}{{{\sigma }_{V}}} \\ 
 & =\frac{\sum\limits_{i \in S}{\left({{A}_{i}}(\lambda )-\overline{A}\right) \left( {{V}_{i}}-\overline{V} \right)}}{\sum\limits_{i \in S}{{{\left( {{V}_{i}}-\overline{V} \right)}^{2}}}}\simeq\frac{{{n}_{c}}\cdot \int\limits_{{{T}_{c}}}{\cos ({{n}_{c}}\cdot (t+\lambda ))\cdot \sin (n{}_{c}\cdot t)dt}}{\int\limits_{{{T}_{c}}}{{{\sin }^{2}}({n}_{c}\cdot t)dt}}=-{{n}_{c}}\cdot \sin \left( {{n}_{c}}\cdot \lambda  \right) \\ 
\end{aligned}.
\end{equation}

Last, we get the \(\chi (\lambda )\):

\begin{equation}\label{25}
\chi (\lambda )=\frac{{{b}_{a-v}}(\lambda )}{{{r}_{a-v}}(\lambda )}=\frac{{{\sigma }_{A(\lambda )}}}{{{\sigma }_{V}}}={{n}_{c}}.
\end{equation}

According to Eq.\ref{25}, it is evident that by temporally shifting the acceleration function and subsequently determining the regression coefficients and correlation coefficients, the construction of the function \(\chi \) allows for the computation of the frequency factor \({{n}_{c}}\) and the frequency of the corresponding speed-time function \({{f}_{c}}=\frac{{{n}_{c}}}{2\pi }\). By introducing \(\chi(\lambda)\), the problem of calculating the frequency of a nonlinear function has been simplified to a linear statistical process.

According to Sec. \ref{subsection3.1}, the linear relationship between NNRD and speed is in accordance with the equation:

\begin{equation}\label{26}
\left\{ \begin{aligned}
  & {{d}_{i}}=b\cdot {{v}_{i}}+{{d}_{0}}+{{\xi }_{i}},\text{  }i=1,2,\cdot \cdot \cdot S \\ 
 & \xi \sim N(0,{{\sigma }^{2}})
\end{aligned} \right..
\end{equation}

Due to the potential anticipation or reaction behavior are observed in the speed and spatial variation of pedestrians \citep{wang2024,tavana2024novel}. Consequently, a time shift function concerning the NNRD can be formulated:

\begin{equation}\label{27}
D=d\left( t+{{\delta }_{A}} \right)={{\mu }_{c2}}\cdot \sin \left( {{n}_{c}}\cdot \left( t+{{\delta }_{A}} \right) \right)+{{c}_{2}}.
\end{equation}

Here, \({{\delta }_{A}}\) denotes the TD of pedestrian motion. 
The coefficients corresponding to Eq.\ref{27} can be computed using the following expression:

\begin{equation}\label{28}
\left\{ \begin{aligned}
  & {{\mu }_{c2}}\simeq b\cdot {{\mu }_{c1}} \\ 
 & {{c}_{2}}\simeq b\cdot {{c}_{1}}+{{d}_{0}} 
\end{aligned} \right..
\end{equation}

Ultimately, we can formulate the coefficient function concerning the TD\(({{\delta }_{A}})\) for functions \(D\) and \(V\).

\begin{equation}\label{29}
{{r}_{d-v}}=\frac{Cov\left( D,V \right)}{{{\sigma }_{D}}{{\sigma }_{V}}}=\frac{\sum\limits_{i \in S}{\left({{D}_{i}}-\overline{D}\right) \left( {{V}_{i}}-\overline{V} \right)}}{\sqrt{\sum\limits_{i \in S}{{{\left({{D}_{i}}-\overline{D}\right)}^{2}}\sum\limits_{i \in S}{{{\left( {{V}_{i}}-\overline{V} \right)}^{2}}}}}}\simeq\frac{\int\limits_{{T}_{c}}{\sin \left( {{n}_{c}}\cdot \left( t+{{\delta }_{A}} \right) \right)\cdot \sin \left( {{n}_{c}}\cdot t \right)dt}}{\sqrt{\int\limits_{{T}_{c}}{{{\sin }^{2}}({{n}_{c}}\cdot(t+{{\delta }_{A}}))dt}\cdot \int\limits_{{T}_{c}}{{{\sin }^{2}}(n{}_{c}\cdot t)dt}}}=\cos \left( {{n}_{c}}\cdot {{\delta }_{A}} \right).
\end{equation}

In consideration of the correlation coefficient established between speed samples and NNRD samples, and by Eq.\ref{25} for the derivation of the coefficient \({{n}_{c}}\), the absolute value of TD associated with pedestrian motion can be calculated by Eq.\ref{30}.

\begin{equation}\label{30}
\left| {{\delta }_{A}} \right|=\frac{\arccos \left( {{r}_{d-v}} \right)}{{{n}_{c}}}.
\end{equation}

We have designated the algorithm as CosIn-2 and Algo. \ref{CosIn-2} provided its pseudocode representation. The limitation of the CosIn-2 algorithm lies in its ability to only obtain the absolute value of the TD (\(\left| {{\delta }_{A}} \right|\)), thereby allowing for the evaluation of the TD scale but rendering it incapable of assessing the anticipation and reaction behavioral patterns.

\begin{algorithm}
  \caption{CosIn-2 Algorithm}\label{CosIn-2}
  \SetAlgoLined
  \KwData{Pedestrians $i$ in the ROI $\bm{\lambda}$: $i=1\cdot \cdot \cdot m$ \\
          Individual sample size: $k$ \\
          Total sample size: $S = m \cdot k$ \\
speed-time and NNRD-time data for each pedestrian: \({{D}_{i}}\) And \({{V}_{i}}\) \\
          
  \textbf{Input:}  $D_i = \{ d(t_1), d(t_2), \ldots, d(t_{k}) \}, i=1\cdot \cdot \cdot m$ \\
  \text{  }{  }{  }{  }{  }{  }{  }{  }{  }{  } $V_i = \{ v(t_1), v(t_2), \ldots, v(t_{k}) \}, i=1\cdot \cdot \cdot m$}
  \textbf{Output:} $\left| {{\delta }_{A}} \right|$\\
  \textbf{Step 1:} Construct sine function for speed\\
  $V(t) = \mu_{c1} \cdot \sin(n_c \cdot t) + c_1$\\
  Determine time shift parameter: $\lambda$\\

  \textbf{Step 2:} Calculate acceleration data:\\
  \ForEach { $i$ }
  {
    $A_i(\lambda) = \{ a(t_1 + \lambda), a(t_2 + \lambda), \ldots, a(t_{k} + \lambda) \}$\\
  }

  \textbf{Step 3:} Calculate regression coefficients and correlation coefficients:\\
  See Appx.\ref{Appendix D} and Eqs. \ref{23} and \ref{24} to calculate $b_{a-v}(\lambda)$ and $r_{a-v}(\lambda)$ based on samples $V$ and $A(\lambda)$\\

  \textbf{Step 4:} Calculate frequency factor $n_c$:\\
  $\chi(\lambda) = \frac{b_{a-v}(\lambda)}{r_{a-v}(\lambda)} = \frac{\sigma_{A(\lambda)}}{\sigma_V} = n_c$\\

  \textbf{Step 5:} Calculate correlation coefficient $r_{d-v}$:\\
  See Appx.\ref{Appendix D}  and Eq.\ref{29}  to calculate $r_{d-v}$ based on samples $V$ and $D$\\

  \textbf{Step 6:} Calculate estimated absolute value of TD: $\left| {{\delta }_{A}} \right| $\\
  $\left| {{\delta }_{A}} \right| = \frac{\arccos(r_{d-v})}{n_c}$\\

\end{algorithm}

\subsection{Case Study} \label{subsection5.2}

In this section, the evaluation of the CosIn-2 algorithm is validated through experiment data. Firstly, in Sec. \ref{subsubsection5.2.1}, the CosIn-2 algorithm is applied to calculate the data of single-file motion, as mentioned above. The discrete cross-correlation was utilized as the reference method to evaluate the relative errors and effectiveness of the CosIn-1 and CosIn-2 algorithms. Subsequently, in the following Sec. \ref{subsubsection5.2.2}, the CosIn-2 algorithm is extended to a broader range of instances, employing it in the calculation of TD in crowd cross experiments. This further contributes to the assessment of the algorithm's adaptability.

In statistical analysis, the time shift parameter of function \(A\) is designated as \(\lambda\)=0.2 s. In computational procedures, forward and reverse temporal shifts are executed utilizing time shift parameters. These operations are conducted to compute the mean, thereby ameliorating statistical errors. Additionally, in this section, we set the sampling frequency at 2.5 Hz. According to the approximate calculation strategy of the CosIn-2 algorithm, excessively high sampling frequencies will not significantly improve algorithm precision.

\subsubsection{Single-file Experiment} \label{subsubsection5.2.1}

\begin{figure}[ht!]
\centering
\includegraphics[scale=0.41]{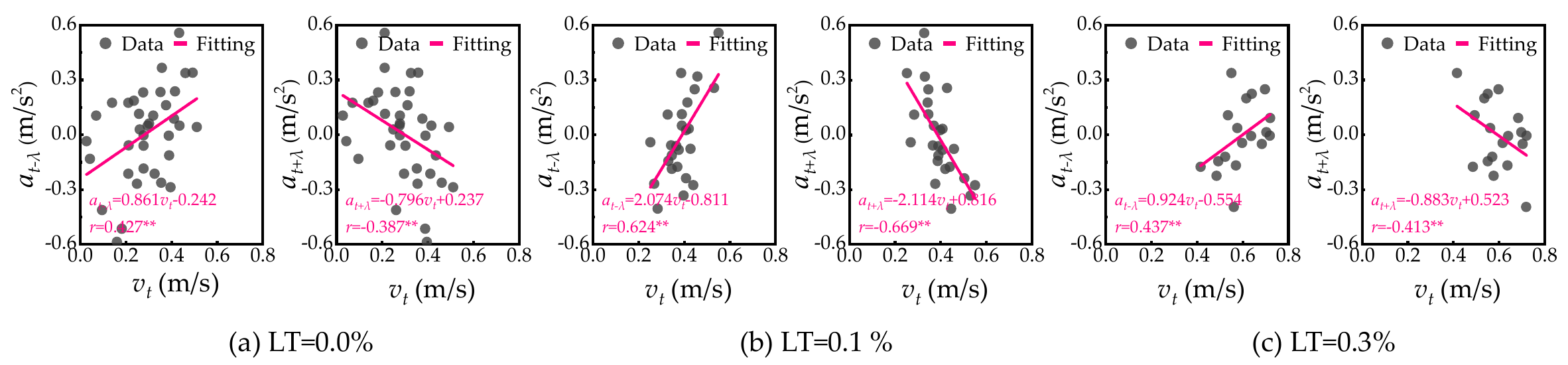}
\captionsetup{singlelinecheck=false, justification=centering}
\caption{The regression and correlation coefficients corresponding to speed and acceleration time shift data under single-file motion contexts. Due to the relatively short sampling duration (all three experimental groups being less than 20 s, sampling frequency: 2.5 Hz). Consequently, the resultant dataset is characterized by a notably limited volume. The significance of the correlation coefficient statistical results is depicted in the figure (*: \(t\)-test, \(p\)<0.05, **: \(t\)-test, \(p\)<0.01).}
\label{fig12}
\end{figure}

In this section, we employ the CosIn-2 algorithm to compute the TD in the context of single-file motion. First, in accordance with Eq.\ref{25}, we perform a statistical analysis on the speed and acceleration data. Fig. \ref{fig12} presents the scatter plot of the velocity and time-shifted acceleration data, along with the corresponding regression coefficients and correlation coefficients.

Based on these results, the frequency factors \(({n}_{c})\) can be derived. Tab. \ref{table2} presents the frequency factors ($n_c$) corresponding to the velocity waveforms for each experimental set, along with the correlation coefficient between the headway and the speed waveforms. In accordance with Eq. \ref{30}, the corresponding absolute value of TD can be calculated, also documented in Tab. \ref{table2}.

\begin{table}[ht]
    \centering
    \caption{Calculation results of CosIn-2.}
    \begin{tabular}{>{\centering\arraybackslash}m{3cm}*{3}{>{\centering\arraybackslash}m{2.5cm}}}
        \toprule
        \textbf{Parameters} & \textbf{LT=0.0\%} & \textbf{LT=0.1\%} & \textbf{LT=0.3\%} \\
        \midrule
        ${n}_{c}$ & 2.03662 & 3.24183 & 2.12622 \\
        $r_{h-v}$ & 0.50426 & 0.11357 & 0.47291 \\
        $\left| {{\delta }_{A}} \right|$ (s) & 0.511765 & 0.449431 & 0.507099 \\
        \bottomrule
    \end{tabular}
\label{table2}
\end{table}

A comparative analysis was conducted using the discrete cross-correlation method as the baseline, based on the results of the CosIn-1 and CosIn-2 algorithms. The computational results for the three datasets are shown in Table 3, which presents the relative errors corresponding to the baseline method. From the data in the table, it can be observed that the relative error of the CosIn-1 algorithm is maintained at a low level, and the longer the sampling duration, the smaller the corresponding relative error. In comparison, the CosIn-2 algorithm exhibits a larger relative error, reaching as high as 59.93\% when LT=0\%.

\begin{figure}[ht!]
\centering
\includegraphics[scale=0.54]{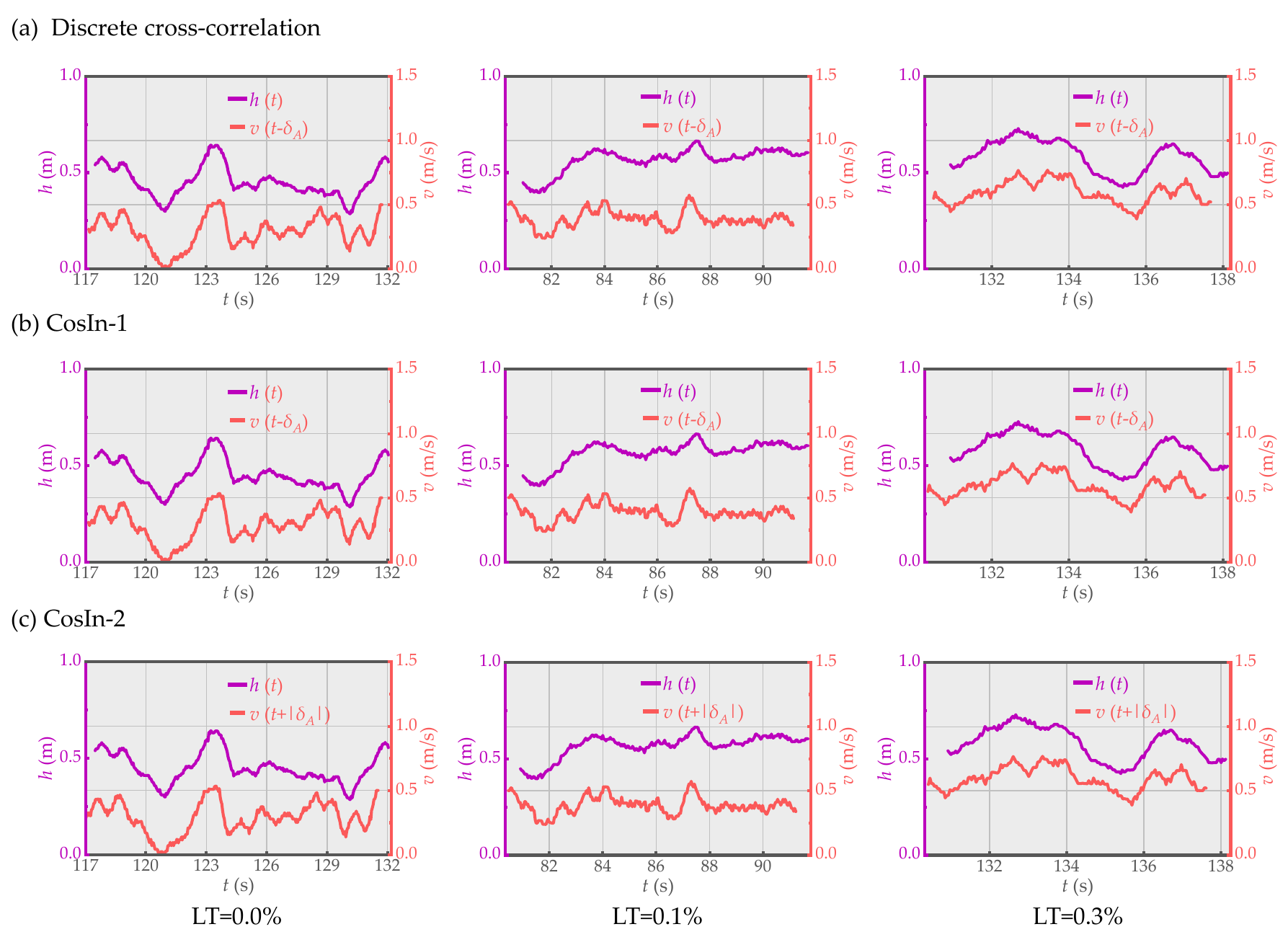}
\captionsetup{singlelinecheck=false, justification=centering}
\caption{
Temporal data of speed profiles and headway profiles for individual pedestrians under the context of specified time shift \({({\delta }_{A})}\). Please note that the subfigures (a) and (b) are not identical, corresponding displacement (\(\delta_{A}\)) is provided in Tab.\ref{table3}.}
\label{fig13}
\end{figure}

Fig. \ref{fig13} illustrates the time-shift compensation curves corresponding to the results obtained from the baseline (discrete cross-correlation), CosIn-1, and CosIn-2 algorithms. As can be seen from the figure, the evaluation results of the three methods do not exhibit significant differences. This is because the TD scales in the three experiments range between 0.3-0.6 seconds, thus, even though the maximum relative error reaches 59.93\% in Table 3, the numerical results remain very close.

\begin{table}[htbp]
  \centering
  \caption{Results and relative error of the CosIn-1 algorithm and CosIn-2 algorithm (sampling durations: 14.56 s for LT = 0\%, 10.76 s for LT = 0.1\%, and 7.20 s for LT = 0.3\%).}
  
\begin{tabular}{>{\centering\arraybackslash}m{5.2cm}*{3}{>{\centering\arraybackslash}m{2.7cm}}}
    \toprule
    \textbf{Algorithms} & \textbf{LT=0.0\%} & \textbf{LT=0.1\%} & \textbf{LT=0.3\%} \\
    \midrule

    Discrete cross-correlation: $\delta_A$ (s)  & -0.320000 & -0.560000 & -0.440000 \\
    CosIn-1: $\delta_A$ (s) (relative error (\%))  & -0.329866(3.08\%) & -0.540323(3.51\%) & -0.590063(34.11\%) \\
    CosIn-2: $|\delta_A|$ (s) (relative error (\%)) & 0.511765(59.93\%) & 0.449431(19.74\%) &0.507099(6.71\%) \\
    \bottomrule
  \end{tabular}
  \label{table3}
\end{table}

Discrete cross-correlation involves calculating the inner product of paired sample points and summing them up to obtain an optimally aligned result through linear translation, with a computational time complexity of \(O({n}^2)\). In comparison, the primary computational cost of the CosIn-1 algorithm comes from the DFT. Utilizing the Fast Fourier Transform (FFT) reduces this complexity to \( O({n} \log {n}) \). The CosIn-2 algorithm adopts an approximation method based on the assumed conditions. While this leads to a reduction in computational accuracy, it significantly simplifies the computational process. The required calculations are limited to basic statistical computations and inverse trigonometric functions. The algorithm's complexity increases linearly with the sampling scale, enabling real-time computation of TD. For a comparative analysis between the discrete cross-correlation method, CosIn-1 and CosIn-2 algorithms, see Tab.~\ref{table4}.

\begin{table}[h]
\centering
\caption{
Comparison of the CosIn-1 algorithm and CosIn-2 algorithm, among these, \( n \) represents the number of samples.}

\begin{tabular}{cccc}
\toprule
\textbf{Algorithms} & \textbf{Precision} & \textbf{Calculation} & \textbf{Time Complexity} \\
\midrule
Discrete Cross-correlation & High & Cross-correlation Computation & \(O({n}^{2})\) \\
\midrule
CosIn-1 & High & Fast Fourier Transform (FFT) & \(O({n}\log {n})\) \\

\midrule
CosIn-2 & Medium & Calculation of Regression and Correlation Coefficients & \(O({n})\) \\

\bottomrule
\end{tabular}
\label{table4}
\end{table}

\begin{figure}[ht!]
\centering
\includegraphics[scale=0.4]{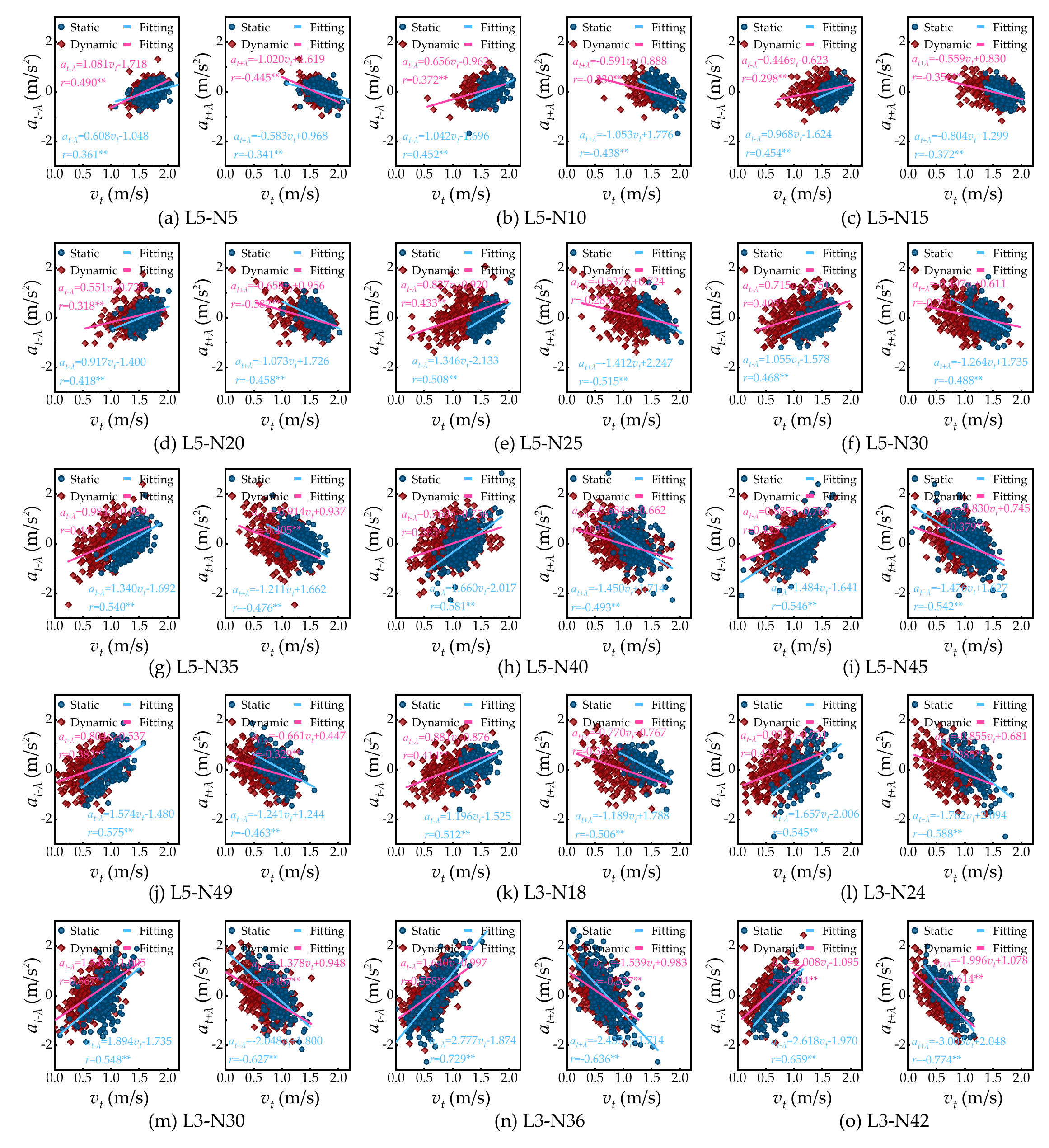}
\caption{The regression and correlation coefficients corresponding to speed and acceleration time shift data under static and dynamic contexts (sampling frequency: 2.5 Hz).The significance of the correlation coefficient statistical results is depicted in the figure (*: \(t\)-test, \(p\)<0.05, **: \(t\)-test, \(p\)<0.01).}
\label{fig14}
\end{figure}

\subsubsection{Crowd-cross Experiment} \label{subsubsection5.2.2}

In this section, we undertake a case validation of the CosIn-2 algorithm based on a crowd-cross experiment. The experimental scenarios are illustrated in Appx.\ref{Appendix A}.

The experiment is designed as a reference pedestrian crossing a fixed experiment area. By varying the number of participants, differential global densities are obtained. Reference participants cross the experimental area, and repeated observations are conducted over multiple trials. Pedestrians in the experiment area adopt two movement modes: static and dynamic. In the static mode, participants remain stationary, while in the dynamic mode, participants move freely. More specific experimental details can be found in \citet{wang2023exploring}. A total of 30 experimental sets were conducted, encompassing static and dynamic contexts with a global density ranging from  \(0.2\text{ ped}/{{\text{m}}^{2}}-3.45\text{ ped}/{{\text{m}}^{2}}\). Statistical calculations of TD were performed for each set. It should be noted that our observations are not confined to a specific ROI but rather focus on the crossing pedestrians. This does not affect the validity of our evaluation process.

\begin{figure}[ht!]
\centering
\includegraphics[scale=0.7]{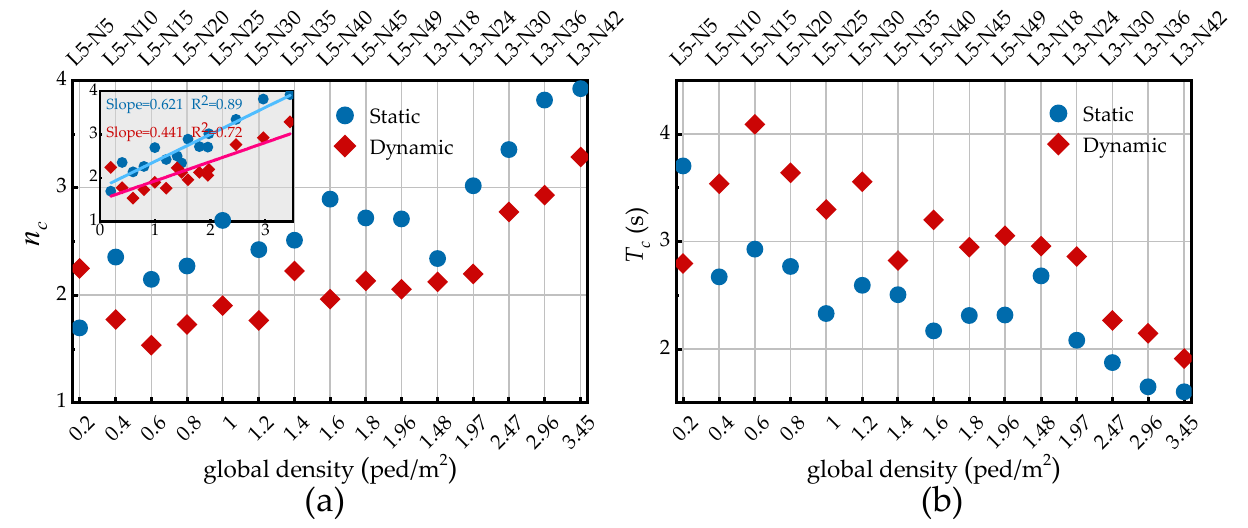}
\caption{The frequency factors and periods corresponding to approximated pedestrian speed functions under varying global densities.}
\label{fig15}
\end{figure}

\begin{figure}[ht!]
\centering
\includegraphics[scale=0.7]{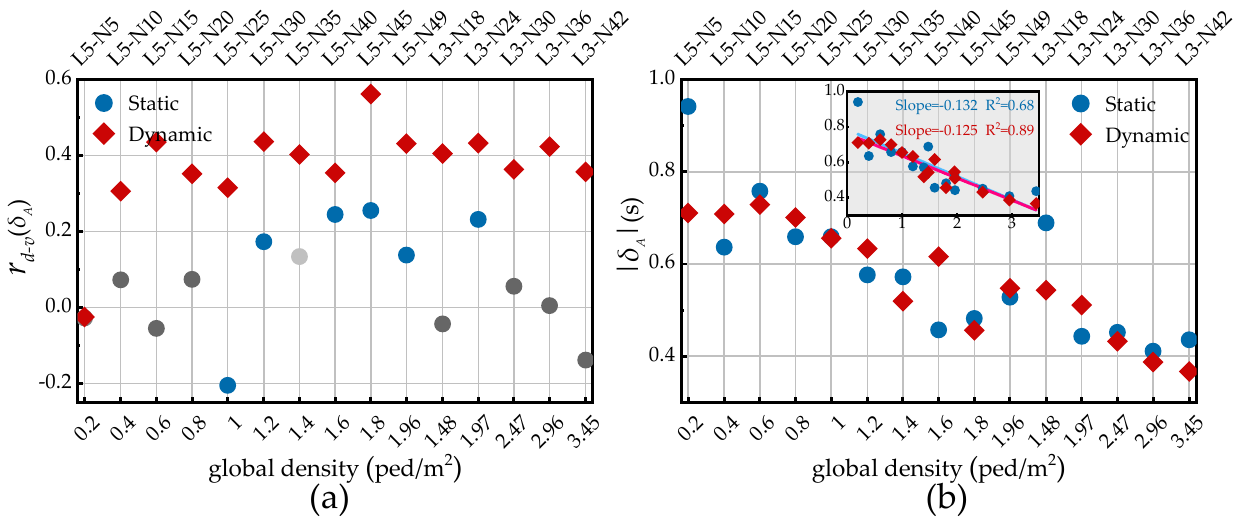}
\caption{Statistical results of the correlation coefficients between pedestrian speed and NNRD \((\phi=\pi/2)\) data under various global densities, along with the corresponding TD ( deep gray symbols indicate situations where the \(p\) > 0.05 in bilateral \(t\)-tests, while shallow gray symbols denote instances where  0.01 < \(p\) < 0.05 in bilateral \(t\)-tests, other colors indicate situations where \(p\) < 0.01 in bilateral \(t\)-tests).}
\label{fig16}
\end{figure}

Based on the data presented in Fig. \ref{fig14}, the frequency factor \(({{n}_{c}})\) and the period \(({{T}_{c}})\) of speed-time functions \((V)\) were computed according to Eqs. \ref{22} and \ref{25}, as illustrated in Fig. \ref{fig15}. The frequency factor of the speed-time functions exhibited a rising trend with an increase in global density, indicating that the higher frequency of pedestrian speed fluctuations at high-density conditions. Moreover, under equivalent global density conditions, the period was found to be shorter when pedestrians cross in the static context. After computing the frequency factor \(({{n}_{c}})\) pertaining to speed in each experiment, we calculated the correlation coefficient \(({{r}_{d-v}})\) between NNRD \((\phi=\pi/2)\) and speed, as illustrated in Fig. \ref{fig16}(a). According to Eq.\ref{30}, the TD \( ({{\delta }_{A}})\) for each experiment were ultimately derived, as depicted in Fig. \ref{fig16}(b).

 From Fig. \ref{fig16}(b), it can be observed that, the TD decreases with the increase in global density (see the subplot of Fig. \ref{fig16}(b)). These results illustrated that, during crossing motion, as the global density increases, the TD scale exhibits a trend of linear contraction. This mechanism is consistent with our conjecture presented in Sec. \ref{subsection2.3}.

\section{Conclusions} \label{section6}

Precise assessment of TD in pedestrian motion is significant  for  identification of pedestrian behavior and crowd pattern. To tackle challenges associated with accurate and rapid TD evaluation, the CosIn algorithm, comprising the CosIn-1 and CosIn-2 algorithms, is introduced in this paper. CosIn-1 computes precise TD values for individual motion, while CosIn-2 approximates TD values for real-time assessment of crowd dynamics.

The CosIn-1 algorithm achieves the analytically computation of TD through Fourier series analysis. The corresponding Fourier series are employed in the calculation process, leading to the solution of differential-algebraic equations for deriving the precise TD solution. The algorithm demonstrates heightened precision, with computational error approaching zero when the sampling size is extensive enough to warrant a sufficiently high order of the Fourier series.

The CosIn-2 algorithm estimates TD from a statistical perspective based on the assumption that the pedestrian speed-time function follows a sine (or cosine) pattern. Through statistical calculations, an approximate solution for TD within the sample data can be ascertained. The time complexity of the CosIn-2 algorithm increases linearly with the augmentation of sample size, enabling real-time TD evaluation within crowds.

In the case study analysis, the discrete cross-correlation method was employed as a baseline to validate the adaptability and computational advantages of the CosIn-1 and CosIn-2 algorithms. Statistical analysis of the TD scale in the crowd-crossing experiment indicates that as global density increases, the pedestrian TD scale tends to shrink linearly. This observation suggests that under extreme conditions (\(\delta \to 0\), \(\tau \to 0\)), the particle model can be utilized for simplified analysis of the crowd.

This paper introduces a new perspective and methodology for pedestrians' behavior patterns assessment and crowd risk management. The CosIn-1 algorithm developed herein demonstrates the capacity to address virtually all instances of temporal delay phenomena in natural signals, extending its applicability across diverse domains. Further validation is requisite to substantiate the efficacy of this algorithm.

\centerline{}
\section*{Data Availability}
The experimental data can be found here: {\url{https://doi.org/10.34735/ped.2019.4}} (Pedestrian Dynamics Data Archive) and {\url{https://drive.google.com/drive/folders/1NYVnRp0z8VPuskfezMr51gB-sraOf6Iq?usp=drive_link}} (Google Drive, includes raw data, graphical files, and codes).

\centerline{}
\section*{Acknowledgments}
We would like to thank Hisashi Murakami, Francesco Zanlungo, and Claudio Feliciani for the open-source data, and Xiao Yao for providing the experimental data. This work was supported by the National Natural Science Foundation of China (Grant No. 52072286, 71871189, 51604204), and the Fundamental Research Funds for the Central Universities (Grant No. 2022IVA108).

\appendix

\section{Experimental Scenario Illustration} \label{Appendix A}
\begin{figure}[ht!]
\centering
\includegraphics[scale=0.18]{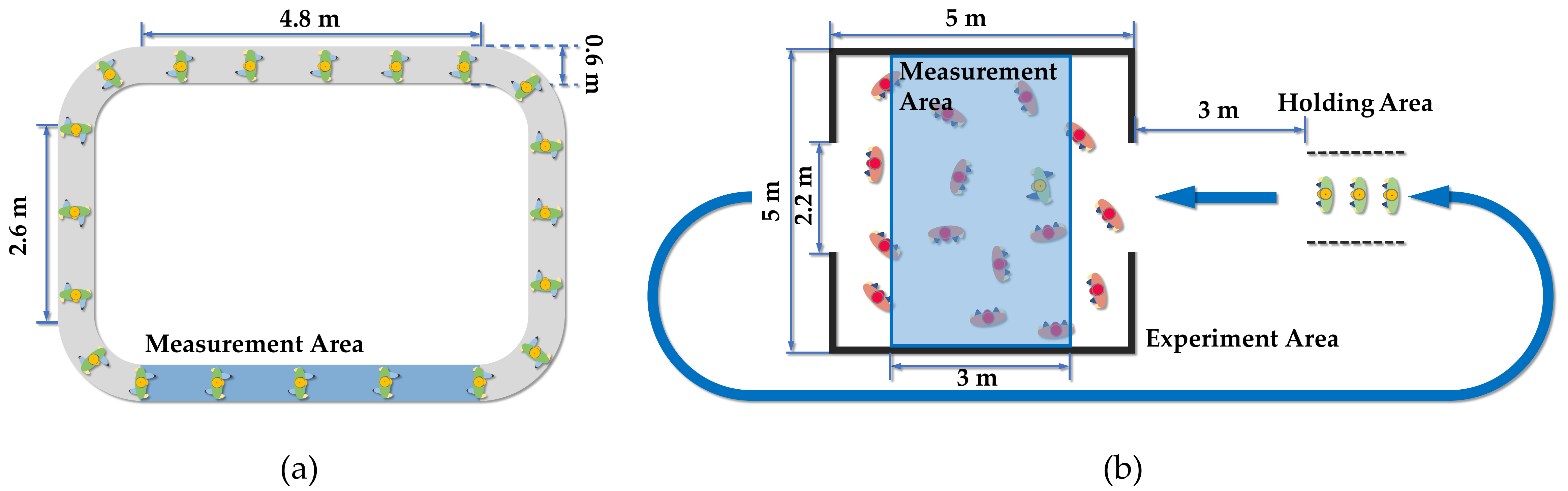}
\caption{(a) Illustrative diagram of  single-file experiment, (b) Illustrative diagram of crowd cross experiment (low-density experiment). }
\label{fig17}
\end{figure}

The experimental scenarios of single-file motion and crowd cross motion in Fig. \ref{fig17} are depicted. Detailed experimental configurations and procedural specifics can be found in the works of \citet{cao2019dynamic}, and \citet{wang2023exploring}, as referenced.

\section{TTC Calculation for Pedestrians in 2D Plane} \label{Appendix B}

\begin{figure}[ht!]
\centering
\includegraphics[scale=1.2]{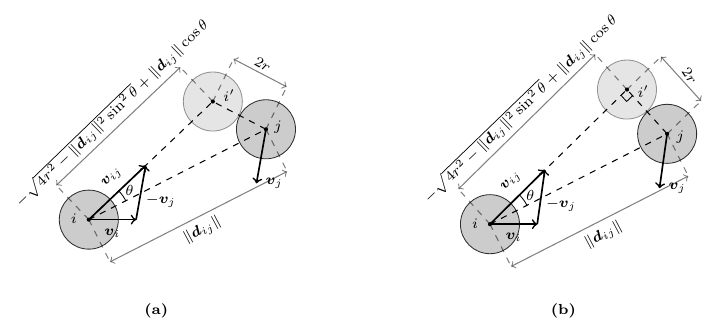}
\caption{Schematic of the pedestrian collision process: (a) collision process between pedestrians \(i\) and \(j\), (b) critical condition for the collision between pedestrians \(i\) and \(j\).}
\label{fig18}
\end{figure}

In two-dimensional motion, pedestrians can be approximated as disks with a radius of \( r \) for simplification purposes. Under this assumption, the TTC between pedestrian \( i \) and their nearest neighbor, pedestrian \( j \), can be determined based on their relative distance and velocities, as depicted in Fig. \ref{fig18}(a). The velocities of pedestrians \( i \) and \( j \) are denoted by \( \bm{v}_i \) and \( \bm{v}_j \), respectively. Here, \( \bm{d}_{ij} \) and \( \bm{v}_{ij} \) represent the relative distance and relative velocity between pedestrian \( i \) and pedestrian \( j \) in the HFA (where \( \phi = \pi \)), respectively. The TTC is computed by projecting the relative velocity vector along the path where the future position of pedestrian \( i \) (denoted as \( i' \)) would result in a collision with pedestrian \( j \), formulated as:

\begin{equation}\label{31}
\tau = \frac{-\sqrt{4r^2 - \|\bm{d}_{ij}\|^2 \sin^2 \theta} + \|\bm{d}_{ij}\| \cos \theta}{\|\bm{v}_{i,j}\|}.
\end{equation}

Where \( \theta \) denotes the angle between \( \bm{d}_{ij} \) and \( \bm{v}_{ij} \). If this angle is sufficiently large, it can be ensured that pedestrians \( i \) and \( j \) will never collide. The critical angle, as illustrated in Fig. \ref{fig18}(b), corresponds to the scenario where  \( i' j \) is orthogonal to \( i' i \) at the moment of collision, expressed as \(\theta = \arcsin(2r / \|\bm{d}_{ij}\|) \).

\section{Cross-Correlation} \label{Appendix C}

Cross-correlation is a quantitative measure of the similarity between two series as a function of the time shift (i.e. TD) of one relative to the other. Upon computing the cross-correlation between the two signals, the maximum (or minimum, in the case of negatively correlated signals) of the cross-correlation function signifies the point in time at which the signals are optimally aligned \citep{lee1949application,bracewell1966fourier}, calculated as:

\begin{equation}\label{32}
\delta_A = \arg \max_{t \in \mathbb{R}} \left( R_{fg}( \delta) \right),
\end{equation}

The cross-correlation $R_{fg}(\delta)$ of two periodic signals $f(t)$ and $g(t)$ with common period $T$ is defined as:

\begin{equation}\label{33}
R_{fg}(\delta) = (f \otimes g)(\delta) = \int\limits_{t \in T} f(t) \cdot g(t + \delta) \, dt \simeq \sum_{t \in T} f(t) \cdot g(t + \delta)
\end{equation}

Here, \(\otimes\) represents the cross-correlation operator, and its discrete form, as shown in Eq.\ref{33}, corresponds to the calculation of discrete cross-correlation.

Accordingly, the Zero-Normalized Cross-Correlation (\(ZNCC\)) between two periodic signals $f(t)$ and $g(t)$ over a common period $T$ at a delay $\delta$ is defined as:

\begin{equation}\label{34}
ZNCC(\delta) = \frac{\int\limits_{t \in T} (f(t) - \overline{f})(g(t+\delta) - \overline{g}) \, dt}{\sqrt{\int\limits_{t \in T} (f(t) - \overline{f})^2 \, dt} \sqrt{\int\limits_{t \in T} (g(t+\delta) - \overline{g})^2 \, dt}}.
\end{equation}

There, the numerator is the dot product of the centered signals at the time delay $\delta$, and the denominator is the product of the norms of the centered signals, ensuring that the correlation measure is normalized to the range \([-1, 1]\). When \(\delta=0\), \(ZNCC(0)\) equals the pearsons' correlation coefficient.

\section{The Sample Statistical Relationship Between Functions \(X\) and \(Y\)} \label{Appendix D}

Assuming a linear relationship exists among corresponding samples derived from functions \(X\) and \(Y\), expressed as:

\begin{equation}\label{35}
{{y}_{i}}=b\cdot {{x}_{i}}+c+{{\xi }_{i}},\text{  } i=1,2,\cdot \cdot \cdot k.
\end{equation}

In the given expression, \(b\) represents the regression coefficients of the sample, \(c\) denotes the constant term, and \(\xi \) signifies the error term. According to the Central Limit Theorem, it is evident that \(\xi \sim N(0,{{\sigma }^{2}})\).

In light of this, we can derive:

sample mean:

\begin{equation}\label{36}
\bar{x}=\frac{\sum\limits_{i=1}^{k}{x{}_{i}}}{k},\bar{y}=\frac{\sum\limits_{i=1}^{k}{y{}_{i}}}{k},
\end{equation}

sample standard deviation:

\begin{equation}\label{37}
{{\sigma }_{x}}=\sqrt{\frac{\sum\limits_{i=1}^{k}{{{({{x}_{i}}-\overline{x})}^{2}}}}{k-1}},{{\sigma }_{y}}=\sqrt{\frac{\sum\limits_{i=1}^{k}{{{({{y}_{i}}-\overline{y})}^{2}}}}{k-1}},
\end{equation}

sample covariance:
\begin{equation}\label{38}
Cov(x,y)=\frac{\sum\limits_{i=1}^{k}{({{x}_{i}}-\overline{x})({{y}_{i}}-\overline{y})}}{k-1},
\end{equation}

sample correlation coefficient:

\begin{equation}\label{39}
r=\frac{Cov(x,y)}{{{\sigma }_{x}}{{\sigma }_{y}}}=\frac{\sum\limits_{i=1}^{k}{({{x}_{i}}-\overline{x})(y-\overline{y})}}{\sqrt{\sum\limits_{i=1}^{k}{{{(x-\overline{x})}^{2}}\sum\limits_{i=1}^{k}{{{(y-\overline{y})}^{2}}}}}},
\end{equation}

sample regression coefficient:

\begin{equation}\label{40}
b=r\cdot \frac{{{\sigma }_{y}}}{{{\sigma }_{x}}}=\frac{\sum\limits_{i=1}^{k}{({{x}_{i}}-\overline{x})({{y}_{i}}-\overline{y})}}{\sum\limits_{i=1}^{k}{{{({{x}_{i}}-\overline{x})}^{2}}}}.
\end{equation}

\section{Fourier Coefficients} \label{Appendix E}

In Sec. \ref{subsection4.2}, the Fourier coefficients corresponding to the expansions of the pedestrian speed-time and headway-time functions in the three experimental sets are provided in Tab.\ref{table5} through Tab.\ref{table7}.

\begin{table}
\centering
\tiny
\caption{Coefficients Table (LT=0.0\%).}
\begin{tabular}{cccccccccc}
\toprule
\multirow{4}{*}{Speed} & \multirow{2}{*}{$\alpha$} & $\alpha_0$ & $\alpha_1$ & $\alpha_2$ & $\alpha_3$ & $\alpha_4$ & $\alpha_5$ & $\alpha_6$ & $\alpha_7$ \\
\cmidrule(lr){3-10}
& & 0.290915068 & 0.02284452 & 0.055571709 & -0.026123571 & -0.037715758 & 0.055616353 & -0.011674572 & 0.020321874 \\
\midrule
& \multirow{2}{*}{$\beta$} & & $\beta_1$ & $\beta_2$ & $\beta_3$ & $\beta_4$ & $\beta_5$ & $\beta_6$ & $\beta_7$ \\
\cmidrule(lr){4-10}
& & & -0.054503196 & 0.003130453 & 0.109661315 & -0.014971023 & 0.016124969 & -0.025325842 & 0.000814358 \\
\midrule
\multirow{4}{*}{Headway} & \multirow{2}{*}{$\mu$} & $\mu_0$ & $\mu_1$ & $\mu_2$ & $\mu_3$ & $\mu_4$ & $\mu_5$ & $\mu_6$ & $\mu_7$ \\
\cmidrule(lr){3-10}
& & 0.451810959 & -0.008414193 & 0.066565337 & 0.046247833 & -0.015579571 & 0.02947325 & -0.010387931 & 0.010159479 \\
\midrule
& \multirow{2}{*}{$\eta$} & & $\eta_1$ & $\eta_2$ & $\eta_3$ & $\eta_4$ & $\eta_5$ & $\eta_6$ & $\eta_7$ \\
\cmidrule(lr){4-10}
& & & 0.027670199 & -0.002484081 & 0.055557336 & -0.002624653 & -0.010982733 & -0.01178945 & -0.009699807 \\
\midrule
$\alpha_8$ & $\alpha_9$ & $\alpha_{10}$ & $\alpha_{11}$ & $\alpha_{12}$ & $\alpha_{13}$ & $\alpha_{14}$ & $\alpha_{15}$ & $\alpha_{16}$ & $\alpha_{17}$ \\
\midrule
-0.015209043 & 0.011584708 & 0.006445499 & -0.000385228 & -0.008292588 & 0.02126092 & 4.56E-05 & 0.012907615 & 0.000113251 & 0.00310148 \\
\midrule
$\beta_8$ & $\beta_9$ & $\beta_{10}$ & $\beta_{11}$ & $\beta_{12}$ & $\beta_{13}$ & $\beta_{14}$ & $\beta_{15}$ & $\beta_{16}$ & $\beta_{17}$ \\
\midrule
0.001620904 & 1.32E-05 & 0.017201901 & 0.004051416 & 0.019368708 & -0.019597586 & -0.01514036 & -0.027853362 & -0.015438743 & -0.01132739 \\
\midrule
$\mu_8$ & $\mu_9$ & $\mu_{10}$ & $\mu_{11}$ & $\mu_{12}$ & $\mu_{13}$ & $\mu_{14}$ & $\mu_{15}$ & $\mu_{16}$ & $\mu_{17}$ \\
\midrule
-0.006527005 & -0.003352991 & 0.011963474 & 0.005568592 & 0.006803013 & -0.006036283 & -0.004178502 & -0.006210715 & -0.001956469 & -0.003708619 \\
\midrule
$\eta_8$ & $\eta_9$ & $\eta_{10}$ & $\eta_{11}$ & $\eta_{12}$ & $\eta_{13}$ & $\eta_{14}$ & $\eta_{15}$ & $\eta_{16}$ & $\eta_{17}$ \\
\midrule
0.00751552 & -0.008203948 & 0.005806749 & 0.00390651 & 0.00342457 & -0.007065918 & 0.000126778 & -0.003976903 & 0.002226276 & -0.00095966 \\
\midrule
$\alpha_{18}$ & $\alpha_{19}$ & $\alpha_{20}$ & $\alpha_{21}$ & $\alpha_{22}$ & $\alpha_{23}$ & $\alpha_{24}$ & $\alpha_{25}$ & $\alpha_{26}$ & $\alpha_{27}$ \\
\midrule
-0.002620122 & 0.003435405 & 0.005931252 & -0.000659671 & 2.98E-03 & -0.00148256 & 0.004868789 & -0.00067869 & -0.000728144 & 0.000866236 \\
\midrule
$\beta_{18}$ & $\beta_{19}$ & $\beta_{20}$ & $\beta_{21}$ & $\beta_{22}$ & $\beta_{23}$ & $\beta_{24}$ & $\beta_{25}$ & $\beta_{26}$ & $\beta_{27}$ \\
\midrule
0.000811577 & -0.005236766 & -0.005939306 & -0.00799589 & 0.00041109 & -0.00329625 & -0.004042171 & -0.005099723 & -0.002974937 & 0.00058392 \\
\midrule
$\mu_{18}$ & $\mu_{19}$ & $\mu_{20}$ & $\mu_{21}$ & $\mu_{22}$ & $\mu_{23}$ & $\mu_{24}$ & $\mu_{25}$ & $\mu_{26}$ & $\mu_{27}$ \\
\midrule
0.001087023 & -0.000288744 & -0.001936641 & -1.86E-03 & 0.00078444 & 0.000420826 & -0.002490689 & -2.79E-05 & -2.81E-05 & 0.000104719 \\
\midrule
$\eta_{18}$ & $\eta_{19}$ & $\eta_{20}$ & $\eta_{21}$ & $\eta_{22}$ & $\eta_{23}$ & $\eta_{24}$ & $\eta_{25}$ & $\eta_{26}$ & $\eta_{27}$ \\
\midrule
-0.001643148 & -0.002901791 & -0.001531133 & -0.000995903 & -0.000976954 & 0.000852146 & -0.000774913 & 0.000301509 & -0.000411539 & -5.56E-05 \\
\midrule
$\alpha_{28}$ & $\alpha_{29}$ & $\alpha_{30}$ & $\alpha_{31}$ & $\alpha_{32}$ & $\alpha_{33}$ & $\alpha_{34}$ & $\alpha_{35}$ & $\alpha_{36}$ & $\alpha_{37}$ \\
\midrule
0.003120018 & 0.00053766 & 0.001943159 & 0.001275338 & 0.000476877 & -0.000603862 & 0.001278922 & 0.001001159 & -0.000233957 & 0.000245116 \\
\midrule
$\beta_{28}$ & $\beta_{29}$ & $\beta_{30}$ & $\beta_{31}$ & $\beta_{32}$ & $\beta_{33}$ & $\beta_{34}$ & $\beta_{35}$ & $\beta_{36}$ & $\beta_{37}$ \\
\midrule
-0.00110581 & -0.001678369 & -0.00280827 & -0.002624193 & -0.001833082 & -0.002762572 & -0.002183767 & -0.00097821 & -0.001534714 & -0.002411407 \\
\midrule
$\mu_{28}$ & $\mu_{29}$ & $\mu_{30}$ & $\mu_{31}$ & $\mu_{32}$ & $\mu_{33}$ & $\mu_{34}$ & $\mu_{35}$ & $\mu_{36}$ & $\mu_{37}$ \\
\midrule
0.000184396 & -0.000249422 & 0.000489886 & -0.000453146 & -4.26E-04 & 8.62E-05 & -0.000738148 & 0.000387487 & -1.92E-05 & -0.000498391 \\
\midrule
$\eta_{28}$ & $\eta_{29}$ & $\eta_{30}$ & $\eta_{31}$ & $\eta_{32}$ & $\eta_{33}$ & $\eta_{34}$ & $\eta_{35}$ & $\eta_{36}$ & $\eta_{37}$ \\
\midrule
-0.002139519 & 0.000537976 & -0.001153777 & -0.001405519 & -0.000206125 & -0.00016122 & 0.00031602 & -0.001137837 & 1.09E-05 & -0.000910076 \\
\bottomrule
\end{tabular}
\label{table5}
\end{table}

\begin{table}
\centering
\tiny
\caption{Coefficients Table (LT=0.1\%).}
\begin{tabular}{cccccccccc}
\toprule
\multirow{4}{*}{Speed} & \multirow{2}{*}{$\alpha$} & $\alpha_0$ & $\alpha_1$ & $\alpha_2$ & $\alpha_3$ & $\alpha_4$ & $\alpha_5$ & $\alpha_6$ & $\alpha_7$ \\
\cmidrule(lr){3-10}
& & 0.38957037 & -0.019542814 & -0.001824263 & 0.055642963 & 0.008318904 & 0.004187245 & 0.023878894 & -0.025038449 \\
\midrule
& \multirow{2}{*}{$\beta$} & & $\beta_1$ & $\beta_2$ & $\beta_3$ & $\beta_4$ & $\beta_5$ & $\beta_6$ & $\beta_7$ \\
\cmidrule(lr){4-10}
& & & 0.000215405 & -0.025875422 & -0.007537632 & 0.004127824 & 0.026456426 & -0.00650778 & 0.013819983 \\
\midrule
\multirow{4}{*}{Headway} & \multirow{2}{*}{$\mu$} & $\mu_0$ & $\mu_1$ & $\mu_2$ & $\mu_3$ & $\mu_4$ & $\mu_5$ & $\mu_6$ & $\mu_7$ \\
\cmidrule(lr){3-10}
& & 0.566411111 & -0.033380149 & -0.025906771 & -0.006659553 & -0.000765429 & 0.008838003 & 0.001111877 & 0.003292228 \\
\midrule
& \multirow{2}{*}{$\eta$} & & $\eta_1$ & $\eta_2$ & $\eta_3$ & $\eta_4$ & $\eta_5$ & $\eta_6$ & $\eta_7$ \\
\cmidrule(lr){4-10}
& & & -0.027534407 & -0.031712176 & -0.052871702 & -0.011722952 & -0.00784949 & -0.016253697 & 0.001764293 \\
\midrule
$\alpha_8$ & $\alpha_9$ & $\alpha_{10}$ & $\alpha_{11}$ & $\alpha_{12}$ & $\alpha_{13}$ & $\alpha_{14}$ & $\alpha_{15}$ & $\alpha_{16}$ & $\alpha_{17}$ \\
\midrule
0.004812986 & -0.008339249 & -0.012088451 & 0.014858999 & -0.001462313 & -0.007919071 & 0.004176651 & 0.006887949 & -0.000442645 & -0.001162278 \\
\midrule
$\beta_8$ & $\beta_9$ & $\beta_{10}$ & $\beta_{11}$ & $\beta_{12}$ & $\beta_{13}$ & $\beta_{14}$ & $\beta_{15}$ & $\beta_{16}$ & $\beta_{17}$ \\
\midrule
0.015479566 & -0.017819339 & 0.00588417 & 0.000627766 & 0.018413547 & 0.009990497 & -0.004158269 & 0.005678006 & 0.014940266 & -0.003349157 \\
\midrule
$\mu_8$ & $\mu_9$ & $\mu_{10}$ & $\mu_{11}$ & $\mu_{12}$ & $\mu_{13}$ & $\mu_{14}$ & $\mu_{15}$ & $\mu_{16}$ & $\mu_{17}$ \\
\midrule
0.006895207 & -0.004247609 & 0.004796933 & 0.004262642 & 0.003649268 & 0.002551364 & -0.002045371 & 0.000535486 & 0.002792228 & -0.000489345 \\
\midrule
$\eta_8$ & $\eta_9$ & $\eta_{10}$ & $\eta_{11}$ & $\eta_{12}$ & $\eta_{13}$ & $\eta_{14}$ & $\eta_{15}$ & $\eta_{16}$ & $\eta_{17}$ \\
\midrule
-0.011005204 & -0.002348306 & -0.003997139 & -0.006013818 & -0.004425298 & -0.002424553 & -0.004390269 & -0.002647123 & -0.004594184 & -0.003209016 \\
\midrule
$\alpha_{18}$ & $\alpha_{19}$ & $\alpha_{20}$ & $\alpha_{21}$ & $\alpha_{22}$ & $\alpha_{23}$ & $\alpha_{24}$ & $\alpha_{25}$ & $\alpha_{26}$ & $\alpha_{27}$ \\
\midrule
-0.00564 & -0.0075 & -0.00302 & 0.000358 & -0.00357 & -0.00142 & -0.00097 & -0.00053 & -0.00092 & -0.00078 \\
\midrule
$\beta_{18}$ & $\beta_{19}$ & $\beta_{20}$ & $\beta_{21}$ & $\beta_{22}$ & $\beta_{23}$ & $\beta_{24}$ & $\beta_{25}$ & $\beta_{26}$ & $\beta_{27}$ \\
\midrule
0.009513 & 0.004453 & 0.002663 & 0.001611 & 0.003072 & 0.003167 & 0.003117 & 0.000107 & 0.00133 & 0.001799 \\
\midrule
$\mu_{18}$ & $\mu_{19}$ & $\mu_{20}$ & $\mu_{21}$ & $\mu_{22}$ & $\mu_{23}$ & $\mu_{24}$ & $\mu_{25}$ & $\mu_{26}$ & $\mu_{27}$ \\
\midrule
0.002169 & 0.000428 & 0.001214 & 0.000489 & 0.001278 & 0.001507 & 0.000582 & -0.00014 & 0.0007 & 0.001825 \\
\midrule
$\eta_{18}$ & $\eta_{19}$ & $\eta_{20}$ & $\eta_{21}$ & $\eta_{22}$ & $\eta_{23}$ & $\eta_{24}$ & $\eta_{25}$ & $\eta_{26}$ & $\eta_{27}$ \\
\midrule
-0.00222 & -0.00245 & -0.00056 & -0.00367 & -0.00087 & -0.00191 & -0.00283 & -0.00239 & -0.00124 & -0.00163 \\
\bottomrule
\end{tabular}
\label{table6}
\end{table}

\begin{table}
\centering
\tiny
\caption{Coefficients Table (LT=0.3\%).}
\begin{tabular}{ccccccccccc}
\toprule
\multirow{4}{*}{Speed} & \multirow{2}{*}{$\alpha$} & $\alpha_0$ & $\alpha_1$ & $\alpha_2$ & $\alpha_3$ & $\alpha_4$ & $\alpha_5$ & $\alpha_6$ & $\alpha_7$ & $\alpha_8$\\
\cmidrule(lr){3-11}
        & & 0.595762431 & -0.047592883 & 0.013987326 & -0.028297225 & -0.010485362 & -0.001286004 & 0.009796053 & -0.003230576 & -0.006366911 \\
\midrule
& \multirow{2}{*}{$\beta$} & & $\beta_1$ & $\beta_2$ & $\beta_3$ & $\beta_4$ & $\beta_5$ & $\beta_6$ & $\beta_7$ & $\beta_8$\\
\cmidrule(lr){4-11}
        & & & 0.057043943 & -0.073073104 & -0.023352397 & -0.000874129 & 0.018265754 & -0.026561135 & 0.010016788 & 0.020381986 \\
\midrule
\multirow{4}{*}{Headway} & \multirow{2}{*}{$\mu$} & $\mu_0$ & $\mu_1$ & $\mu_2$ & $\mu_3$ & $\mu_4$ & $\mu_5$ & $\mu_6$ & $\mu_7$ & $\mu_8$\\
\cmidrule(lr){3-11}
        & & 0.581370166 & 0.006887425 & -0.059551968 & -0.032854978 & 0.00683455 & 0.005878383 & -0.010579501 & 0.005183383 & 0.003876198 \\
\midrule
& \multirow{2}{*}{$\eta$} & & $\eta_1$ & $\eta_2$ & $\eta_3$ & $\eta_4$ & $\eta_5$ & $\eta_6$ & $\eta_7$ & $\eta_8$\\
\cmidrule(lr){4-11}
        & & & 0.07662784 & -0.047389036 & 0.029283671 & 0.00516453 & 0.006017145 & -0.001210552 & 0.003791272 & 0.00412184 \\
\midrule
$\alpha_9$ & $\alpha_{10}$ & $\alpha_{11}$ & $\alpha_{12}$ & $\alpha_{13}$ & $\alpha_{14}$ & $\alpha_{15}$ & $\alpha_{16}$ & $\alpha_{17}$ & $\alpha_{18}$ &  $\alpha_{19}$\\
\midrule
        -0.002990756 & 0.017126504 & -0.003512308 & 0.010638614 & -0.001426619 & -0.002120307 & -0.00368285 & 0.000455486 & -0.000121561 & -0.000454565 & -0.000386929 \\
\midrule
$\beta_9$ & $\beta_{10}$ & $\beta_{11}$ & $\beta_{12}$ & $\beta_{13}$ & $\beta_{14}$ & $\beta_{15}$ & $\beta_{16}$ & $\beta_{17}$ & $\beta_{18}$ & $\beta_{19}$ \\
\midrule
        0.012159644 & 0.012859249 & 0.002398505 & -0.00223711 & 0.005595283 & -0.000784076 & 0.002104884 & 0.002089313 & 0.001096799 & 0.000456679 & 0.000873772 \\
\midrule
$\mu_9$ & $\mu_{10}$ & $\mu_{11}$ & $\mu_{12}$ & $\mu_{13}$ & $\mu_{14}$ & $\mu_{15}$ & $\mu_{16}$ & $\mu_{17}$ & $\mu_{18}$ & $\mu_{19}$\\
\midrule
        0.002353154 & 0.000450083 & -0.000709111 & -0.000368524 & 0.001550585 & -3.93E-05 & 0.001247395 & 0.001101608 & 0.000657442 & -0.002399689 & -0.000612994 \\
\midrule
$\eta_9$ & $\eta_{10}$ & $\eta_{11}$ & $\eta_{12}$ & $\eta_{13}$ & $\eta_{14}$ & $\eta_{15}$ & $\eta_{16}$ & $\eta_{17}$ & $\eta_{18}$ & $\eta_{19}$\\
\midrule
        0.003622926 & -0.002151211 & -0.001007306 & -0.005203496 & 0.000980494 & 0.000686396 & 0.001281215 & 0.000751951 & -0.00012114 & 0.00061134 & 0.001814939 \\
\bottomrule

\end{tabular}

\label{table7}
\end{table}

\bibliographystyle{aasjournal}

\end{document}